\documentclass[bibyear]{aa}  

\usepackage{natbib}
\setcitestyle{notesep={, },aysep={}}
\bibpunct{(}{)}{;}{a}{}{,} 
\usepackage{graphicx}
\usepackage{longtable}
\usepackage{tabularx}
\usepackage{supertabular}
\usepackage{tabularray}
\usepackage{xspace}
\usepackage{booktabs,caption}
\usepackage[flushleft]{threeparttable}
\usepackage{txfonts}
\usepackage{hyperref}

    \hypersetup{colorlinks=true,allcolors=[rgb]{0,0,1}}
\usepackage{amsmath}

\newcommand{\lya}{\mbox{Lyman-$\alpha$}\xspace} 
\newcommand{\xlae}{\mbox{$X_{\textrm{LAE}}$}\xspace}
\newcommand{\ttpaper}{\mbox{\cite{Thai2023}}\xspace}
\newcommand{\mm}{\mbox{$\,\mu\mathrm{m}$}\xspace}
\newcommand{\hypz}{\mbox{\textit{New-Hyperz}}\xspace}
\newcommand{\ewlya}{\mbox{$\mathrm{EW_{Ly\alpha}}$}\xspace}
\newcommand{\sfrlya}{\mbox{$\mathrm{SFR_{Ly\alpha}}$}\xspace}
\newcommand{\sfruv}{\mbox{$\mathrm{SFR_{UV}}$}\xspace}
\newcommand{\fuv}{\mbox{$\mathrm{f_{esc,UV}}$}\xspace}
\newcommand{\flya}{\mbox{$\mathrm{f_{esc,Ly\alpha}}$}\xspace}

\defcitealias{GdlV2019LAELF}{dLV19}
\defcitealias{GdlV2020LAEfrac}{dLV20}
\defcitealias{richard2021atlas}{R21}
\defcitealias{AC2022LLAMAS}{C22}
\begin{document} 
   \title{Evolution of the Lyman-$\alpha$ emitting fraction and UV properties of lensed star-forming galaxies between $2.9<z<6.7$}
 
   \titlerunning{Evolution of the Lyman-$\alpha$ emitting fraction and properties of lensed galaxies between $2.9<z<6.7$}

   \author{I. Goovaerts\inst{1}\and R. Pello\inst{1}\and T. T. Thai\inst{1,2,3}\and P. Tuan-Anh\inst{2,3}\and J. Richard\inst{4}\and A. Claeyssens\inst{5}\and E. Carinos\inst{1}\and G. de la Vieuville\inst{6}\and J. Matthee\inst{7}}

   \institute{Aix Marseille Université, CNRS, CNES, LAM (Laboratoire d’Astrophysique de Marseille), UMR 7326, 13388 Marseille, France\\
              \email{ilias.goovaerts@lam.fr}
         \and     
         Department of Astrophysics, Vietnam National Space Center, Vietnam Academy of Science and Technology, 18 Hoang Quoc Viet, Hanoi, Vietnam
         \and
         Graduate University of Science and Technology, VAST, 18 Hoang Quoc Viet, Cau Giay, Vietnam
         \and
         Univ Lyon, Univ Lyon1, Ens de Lyon, CNRS, Centre de Recherche Astrophysique de Lyon UMR5574, F-69230, Saint-Genis-Laval, France
        \and
        Department of Astronomy, Oskar Klein Centre, Stockholm University, AlbaNova University Centre, SE-106 91 Stockholm, Sweden
         \and
         7 Avenue Cuvier, F-78600, Maisons-Laffitte, France
         \and
         Department of Physics, ETH Zürich, Wolfgang-Pauli-Strasse 27, 8093 Zürich, Switzerland
         }

 
  \abstract
   {Faint galaxies are theorised to have played a major, perhaps dominant role in reionising the Universe. Their properties as well as the \lya emitter fraction, \xlae, could provide useful insight into this epoch.}
   {We use four clusters of galaxies from the Lensed Lyman-alpha MUSE Arcs Sample (LLAMAS) which also have deep HST photometry to select a population of intrinsically faint Lyman Break Galaxies (LBGs) and \lya Emitters (LAEs). We study the interrelation between these two populations, their properties, and the fraction of LBGs that display \lya emission.}
   {The use of lensing clusters allows us to access an intrinsically faint population of galaxies, the largest such sample collected for this purpose: 263 LAEs and 972 LBGs between redshifts of 2.9 and 6.7, \lya luminosities between $39.5\lesssim \mathrm{log(L_{Ly\alpha}) (erg\,s^{-1})}\lesssim42$ and absolute UV magnitudes between $-22\lesssim\mathrm{M_{1500}}\lesssim-12$. As well as matching LAEs and LBGs, we define an LAE+continuum sample for the LAEs which match with a continuum object that is not selected as LBG. Additionally, with the use of MUSE integral field spectroscopy, we detect a population of LAEs completely undetected in the continuum.}
   {We find a redshift evolution of \xlae in line with literature results, with diminished values above $z=6$. This has been taken to signify an increasingly neutral intervening IGM. When inspecting this redshift evolution with different limits on \ewlya and $\mathrm{M_{1500}}$ we find that \xlae for the UV-brighter half of our sample is higher than \xlae for the UV-fainter half, a difference which increases at higher redshift. This is a surprising result and can be interpreted as the presence of a population of low \lya equivalent width (\ewlya), UV-bright galaxies situated in reionised bubbles and overdensities. This result is especially interesting in the context of similar, UV-bright, low \ewlya objects recently detected in and around the epoch of reionisation. We extend to intrinsically fainter objects the previously observed trends of LAEs among LBGs as galaxies with high star-formation rates and low dust content, as well as the strongest LAEs having in general fainter $\mathrm{M_{1500}}$ and steeper UV slopes.}
  {}
   
   \keywords{galaxies: high redshift --
                dark ages, reionisation, first stars --
                gravitational lensing: strong
               }
   
   \maketitle
%

\section{Introduction}
\label{sect:intro}
After the Dark Ages, the neutral gas in the intergalactic medium (IGM) was reionised: the last phase transition undergone by the Universe. This process ended with the hydrogen in the IGM ionised, around $z=6$ \citep{fan2006reionisation,mcgreer2015reionisation,Planck2018reionisation,lu2022subarureionisation}. There are two main candidates thought to be responsible for this process, star-forming galaxies (SFGs) \citep{bouwens2015UVLF, Finkelstein_2015, Livermore_2017} and active galactic nuclei (AGN) \citep{MadauHardtAGN_2015, Grazian2018AGN}. The influence of AGN is likely to be small \citep{Onoue2017AGN, parsa2018noAGNcontribution, Mcgreer2018AGN, JiangAGN2022}. Currently, the favoured candidate is SFGs, particularly faint SFGs \citep{robertson2013reionisation,Bouwens2015planckreionisation,Stark2016review}, although the possibly significant contribution of bright SFGs is still debated \citep{naidu2020brightlya, Mathee2022brightlya}.\\
\indent In order to study these intrinsically faint SFGs, one of the most powerful tools is \lya emission. Galaxies that exhibit \lya emission are called \lya emitters (LAEs). The strength of the \lya line for equivalent widths greater than, for instance, $25\,\AA$, allows us to identify intrinsically faint and/or high redshift galaxies. 
In recent years, this has been exploited from several different angles in order to learn more about galaxies at this epoch as well as the state of the IGM and hence reionisation itself (see, for example, the reviews by \citealt{Stark2016review,dijkstra2016LAEreview,robertson2022Reionisationreview} and references therein).\\
\indent Additionally, the use of lensing allows us to probe fainter galaxies than in blank field surveys, down to \lya luminosities of $\sim10^{39}\,\mathrm{erg\,s^{-1}}$ \citep{bina2016muse,smit2017lensedemissionlines,GdlV2019LAELF,GdlV2020LAEfrac,richard2021atlas,AC2022LLAMAS}, the last four of which are henceforth \citetalias{GdlV2019LAELF}, \citetalias{GdlV2020LAEfrac}, \citetalias{richard2021atlas} and \citetalias{AC2022LLAMAS}. This gives us direct access to the faint populations that are, as mentioned, currently the favoured candidates for the main contributor to reionisation. This avenue has been explored in recent studies with small to medium sample sizes (\citetalias{GdlV2020LAEfrac}; \citealt{fuller2020LAEfrac}). Lensing, however, comes with compromises on sample size and volume of the Universe probed. Even more recently, samples of more significant size (hundreds of objects) have become available, such as the Lensed Lyman-alpha MUSE Arcs Sample (LLAMAS) \citepalias{richard2021atlas,AC2022LLAMAS}. \\
\indent In order to investigate reionisation and the role of SFGs, the fraction of Lyman Break Galaxies (LBGs) that exhibit \lya emission (henceforth referred to as the LAE fraction or $X_{\textrm{LAE}}$) is particularly interesting. The Lyman break is caused by absorption of photons at wavelengths shorter than $912\, \AA$ by neutral hydrogen gas around the galaxy and up to $1216\, \AA$ by the Lyman forest along the line of sight. This can be used to search for galaxies photometrically by using the `drop-out' technique: galaxies will `drop out' of filters bluewards of the Lyman break. \lya emission is scattered by neutral hydrogen in the IGM, ISM and circum-galactic medium (CGM) so whether or not \lya emission is detected from LBGs gives us information about the content of these media, in particular how ionised they are, which has the potential to help in reconstructing the timeline and scale of reionisation (see, for example \citealt{mason2018LAEfrac,arrabal2018LAEfrac,kusakabe2020,leonova2022prevalence,bolan2022neutralfraction}). Studies show a drop in the prevalence of LAEs among LBGs (\xlae) above $z=6$, suggesting the increasing neutrality of the IGM before this time and supporting the established reionisation timeline (\citealt{stark2010keckLAEfrac,stark2011LAEfrac,pentericci2011laefrac/z=7LBG,caruana2014LAEfrac,deBarros2017LAEfraction,pentericci2018LAEfrac,caruana2018,hoag2019reionisation_z=7.7}; \citetalias{GdlV2020LAEfrac}). 
However there are significant uncertainties associated with both the measurement of \xlae and its usage as a probe of the reionisation history. The evolution of \xlae with redshift could also be due to the inherent evolution of one or both of the populations considered, rather than solely as a consequence of the changing state of the IGM \citep{bolton2013IGMlya_alternate,mesinger2015IGMlya}. Progress has been made in understanding the impact of the ISM, CGM and dust attenuation on \lya emission \citep{verhamme2008lyadustsim,zheng2010lya_radiativetransfer,dijkstra2011lyadetectability,kakiichi2016lyamorphology_reionisation} however there is still significant debate on the physics of \lya photon escape, in simulations as well as observations, at different redshifts and how this impacts LAE visibility and hence \xlae \citep{dayal2011lyavisibility,matthee2016calymha_lyaesc,hutter2014lyavisibility,sobral2019predictinglyaesc,smith2022lyatransmission_sim} .\\
\indent Additionally, \xlae has shown some dependence on absolute rest-frame UV magnitude \citep{stark2010keckLAEfrac,stark2011LAEfrac,schaerer2011LAEfrac,schenker2014lyafracfeasibility,kusakabe2020} and significantly on the \lya EW cut above which \xlae is calculated \citep{stark2010keckLAEfrac,caruana2018,kusakabe2020}. It is crucial to be aware of these factors when comparing results in the literature. \\
\indent Equally important are the possible biases introduced by the different methods of selecting both the UV `parent sample' and the LAE sample. When collecting the parent sample using the Lyman break, there is no standard way of performing the selection of these galaxies. Some samples are selected by colour--colour cuts such as \citealt{stark2010keckLAEfrac,pentericci2011laefrac/z=7LBG,bouwens2015UVLF,pentericci2018LAEfrac,bouwens2021UVLF,yoshioka2022LAEfrac}, some by photometric redshifts (\cite{caruana2018,fuller2020LAEfrac,kusakabe2020}, \citetalias{GdlV2020LAEfrac}) (although this probably doesn't have a large effect on the sample as the procedure to find the photometric redshift relies on the Lyman Break in the same way as the colour--colour cuts \citepalias{GdlV2020LAEfrac}). The colour--colour cuts used depend on the instrument and bands used to observe the sample as well as the depth of the observations, leading to different cuts for each study. \\
\indent The exact selection using photometric redshifts is also down to the authors of each individual study, such as the signal-to-noise required for a detection as well as how to deal with the probability distributions provided by most photometric redshift codes. \\
\indent Once the parent sample has been selected, the way in which the search for LAEs is conducted is also not standardised. Some authors select LAEs based on Narrow-Band photometry \citep{arrabal2018LAEfrac,yoshioka2022LAEfrac}, some search for \lya emission among their UV--selected sample using multi-object slit spectroscopy \citep{stark2010keckLAEfrac,stark2011LAEfrac,pentericci2011laefrac/z=7LBG,deBarros2017LAEfraction,pentericci2018LAEfrac,fuller2020LAEfrac} and some use Integral Field Unit (IFU) spectroscopy (\citealt{caruana2018}; \citetalias{GdlV2020LAEfrac};  \citealt{kusakabe2020}). \\
\indent Using slit spectroscopy to search a UV-selected sample for \lya emission can be problematic, as the \lya emission is not always centred on the UV emission, in fact it has often been found to have an offset. \citetalias{AC2022LLAMAS} report a median offset of $\Delta_{\mathrm{Ly\alpha-UV}}=0.66\pm0.14\,\mathrm{kpc}$ and \citet{hoag2019lyaoffset} find an offset corresponding to  $0''.25$ at $z=4.5$. Therefore, slit spectroscopy may not see the \lya emission at all, or may miss some flux from extended emission.\\
\indent The completeness of the different samples is also an important factor to take into account in these studies. Completeness is a correction made to the amount of objects observed to account for those present in the field but not observed. This correction depends on many factors and is different for each study. Several different approaches have been employed in the literature. \citet{stark2010keckLAEfrac} consider the completeness of their \lya detections by inserting fake emission lines across their spectra and attempting to detect them using the original detection process, a method which has been evolved into the complex \lya completeness treatments seen in IFU studies such as \citet{kusakabe2020} and \citetalias{GdlV2020LAEfrac}. \citetalias{GdlV2020LAEfrac} and this study involve the extra complication engendered by lensing fields (see \citetalias{GdlV2020LAEfrac} and \ttpaper). \citetalias{GdlV2020LAEfrac} show that the inclusion of the LAE completeness correction is significant to the calculation of \xlae. \\
\indent The study can be performed on a UV--complete subsample of the LBG population such as in \citet{kusakabe2020} but it is common to assume that one's LBG selection is highly complete for the signal-to-noise requirements imposed in the selection process. This is also an assumption sometimes made for the \lya selection. For studies not involving lensing, one can easily calculate the apparent magnitude at which one becomes incomplete in the UV selection at the 10\% and 50\% level such as in \citet{arrabal2018LAEfrac}.\\
\indent In light of these discrepancies in the different studies, it is not surprising that there is significant disagreement on the precise values and evolution of \xlae further than a general consensus that \xlae rises from lower redshifts to a redshift of $\sim$6, after which it sees a decrease at higher redshifts (see, among others: \citealt{pentericci2011laefrac/z=7LBG,stark2011LAEfrac,deBarros2017LAEfraction,arrabal2018LAEfrac,caruana2018,fuller2020LAEfrac}; \citetalias{GdlV2020LAEfrac}).\\
\indent In addition to the reasons previously mentioned, the scatter in the results from these various studies can come from issues related to sample size, as well as the exact sample used to calculate the fraction. For example, the inclusion limits on $M_{\mathrm{1500}}$ are not homogeneous across all studies, neither are the inclusion limits on the \lya equivalent width. \\
\indent In this paper we investigate faint star-forming galaxies towards the epoch of reionisation ($2.9<z<6.7$), observed behind four lensing clusters in the Hubble Frontier Fields (HFF hereafter; \cite{lotz2017frontier}), specifically chosen for being efficient enhancers of such high-redshift objects. We select LAEs from MUSE IFU spectroscopy and LBGs from deep Hubble photometry and photometric redshifts. We present the largest LAE and LBG combined sample of lensed, intrinsically faint galaxies used for this purpose to date. Our sample reaches as faint as $\mathrm{M_{1500}}\sim-12$ and $\mathrm{log(L_{Ly\alpha}/erg\,s^{-1})\sim39.5}$ after correcting for lensing magnification. This is similar to the latest LBG selection in the HFF from \cite{Bouwens2022UVLF2z9}. Having blindly selected these populations of galaxies, independently of each other, we compare their UV and \lya-derived properties and explore the fraction of LAEs among the LBG population, its redshift evolution and UV magnitude dependence.\\
\indent In Section~\ref{sect:data} we cover the data used from MUSE and the HFF as well as the specific selection criteria we apply for our samples of LAEs and LBGs. We outline the photometric redshifts used and subsequently the blind matching of both populations. Section~\ref{sect:results} outlines our results pertaining to derived properties: equivalent widths, UV slopes and star formation rates, as well as the interrelation between the two populations, including several approaches to the fraction of LAEs among the LBG population. We discuss the implications for reionisation and the properties of these high-redshift galaxies. In Section \ref{sect:conclusion} we summarise our findings and offer perspectives for future surveys.\\
\indent The Hubble constant used throughout this paper is $H_0=70\,\mathrm{km\,s^{-1}Mpc^{-1}}$ and the cosmology: $\Omega_{\Lambda}=0.7$, $\Omega_{\mathrm{m}}=0.3$. All EWs and UV slopes are converted to their rest-frame values and all magnitudes are given in the AB system \citep{OkeGunn1983}. All values of the UV absolute magnitude, defined at $1500\,\AA$ rest frame, $\mathrm{M_{1500}}$, and \lya luminosity are given corrected for magnification. 

\section{Data and Population Selection}
\label{sect:data}
For this study, we combine Multi--Unit Spectroscopic Explorer (MUSE; \cite{Bacon_et_al_2010_SPIE}) IFU observations with the deepest Hubble Space Telescope (HST) photometry available in lensing clusters. The Hubble Frontier Fields clusters \citep{lotz2017frontier} are ideal for this work, specifically, we use Abell 2744, Abell 370, Abell S1063 and MACS 0416 (henceforth A2744,  A370, AS1063 and M0416). The MUSE observations of these clusters are taken from the data collected in \citetalias{richard2021atlas}, from which the LAEs are selected to form the LLAMA Sample \citepalias{AC2022LLAMAS}. The complementary HST data are from the HFF-DeepSpace program (PI: H. Shipley)\footnote{\url{http://cosmos.phy.tufts.edu/~danilo/HFF/Download.html}}.

\begin{table*}
    \caption{Details of each lensing cluster used in this study.}
    $$
    \begin{array}{p{0.07\linewidth}p{0.1\linewidth} p{0.1\linewidth}p{0.06\linewidth}p{0.08\linewidth}p{0.08\linewidth}p{0.05\linewidth}p{0.05\linewidth}p{0.05\linewidth}p{0.07\linewidth}}
        \hline
        \hline
        \noalign{\smallskip}
        Cluster & RA & DEC & Redshift & MUSE Exposure Time (Hrs) & Co-volume (Mpc\textsuperscript{3}) & LAE only & LBG only & LAE + LBG & LAE + continuum\\
        \hline
        \noalign{\smallskip}
        \textbf{A2744} & 00:14:20.702  & -30:24:00.63 & 0.308 & 3.5-7 & 10500 & 24 & 294 & 47 & 50\\
        \noalign{\smallskip}
        \hline
        \noalign{\smallskip}
        \textbf{A370} & 02:29:53.122 & -01:34:56.14 & 0.375 & 1.5-8.5 & 5350 & 8 & 267 & 20 & 14\\
        \noalign{\smallskip}
        \hline
        \noalign{\smallskip}
        \textbf{AS1063} & 22:48:43.975 & -26:05:08.00 & 0.348 & 3.9 & 1970 & 6 & 159 & 7 & 8\\
        \noalign{\smallskip}
        \hline
        \noalign{\smallskip}
        \textbf{M0416S} & 04:16:09.144 & -24:04:02.95 & 0.397 & 11-15 & 1670 & 4 & 81 & 14 & 16\\
        \noalign{\smallskip}
        \hline
        \noalign{\smallskip}
        \textbf{M0416N} & 04:16:09.144 & -24:04:02.95 & 0.397 & 17 & 3420 & 13 & 71 & 12 & 20\\
        \noalign{\smallskip}
        \hline
        \noalign{\smallskip}
        \textbf{Total} &  &  &  &  & 22910 & 55 & 872 & 100 & 108\\
        \noalign{\smallskip}
        \hline
        \hline
        \vspace{0.5cm}
    \end{array}
    $$
    \label{table:data_details}
    \vspace{-5mm}
    \begin{tablenotes}
      \small
      \item \textbf{Notes.} We show the number of each sample group in each cluster (see text). The different exposure times for three of the clusters come from the range of exposure times for different MUSE pointings for these clusters.
    \end{tablenotes}
\end{table*}

\begin{figure*}
    \centering
    \includegraphics[height=8cm]{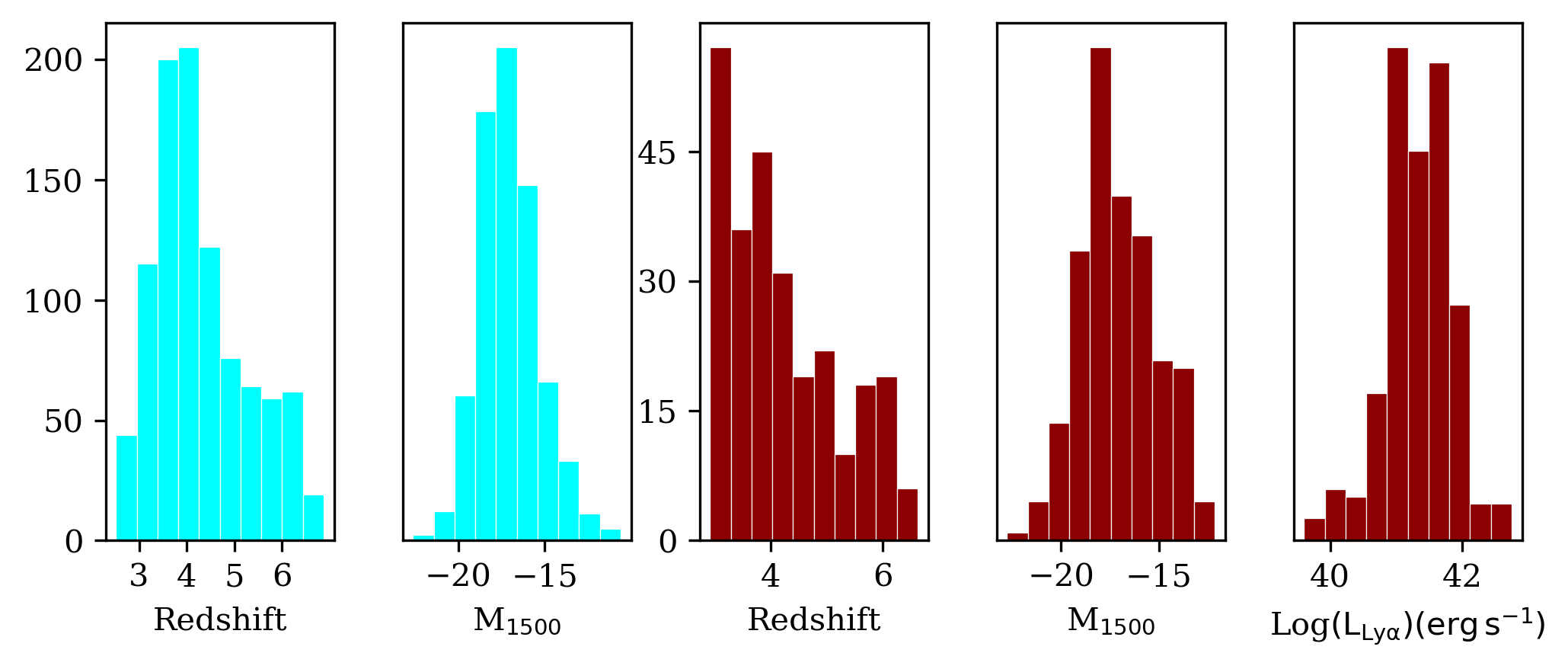}
    \caption{Redshift and $M_{1500}$ (and \lya luminosity for the LAEs) distributions of both samples used in this work; LBGs are shown in blue and LAEs in red.}
    \label{fig:LAELBGzm1500dist}
\end{figure*}

\subsection{MUSE LAEs: LLAMAS}
\label{sect:LAE}

\subsubsection{LAE Selection}
\label{sect:laeselection}
The MUSE data we use for this work is part of the Lensed Lyman-Alpha MUSE Arcs Sample (LLAMAS) \citepalias{AC2022LLAMAS}. These observations were part of the MUSE Guaranteed Time Observing (GTO) program and are comprehensively described in \citetalias{AC2022LLAMAS} as well as \citetalias{richard2021atlas}. The catalogues and lens models for the four clusters used in this work are available online\footnote{\url{https://cral-perso.univ-lyon1.fr/labo/perso/johan.richard/MUSE_data_release/}}. We summarise here the details of the reduced MUSE datacubes and the LAE selection process. \\
The MUSE datacubes contain the flux and variance over a $1\times1\,\mathrm{arcmin}^2$ field of view, with spatial and spectral pixel scales of $0.2''$ and $1.25\,\AA$. The spectral range covers $4750\,\AA$ to $9350\,\AA$, allowing the detection of \lya between redshifts of 2.9 and 6.7. Integration times on the clusters we use for this work vary between 2 and 14 hours, with different pointing configurations (see below). The clusters are all in the range $0.3<z<0.4$, providing magnifications useful to amplify sources in the MUSE \lya redshift range. Details of the four clusters are given in Table~\ref{table:data_details}. Full details covering all 17 LLAMAS clusters can be found in \citetalias{AC2022LLAMAS} and \ttpaper. \\
\indent In order to detect line emission sources such as LAEs, \citetalias{richard2021atlas} follow the prescription laid out in \citet{PW2020MUSEdata}, using the MUSELET software (\citealt{piqueras2019asp})\footnote{\url{https://mpdaf.readthedocs.io/en/latest/muselet.html}}, which is used on MUSE Narrow Band (NB) images. Subsequently, the Source Inspector package \citep{bacon2022musedatareleaseII} is used to identify sources and assign their redshifts. This package allows users to cycle through a list of sources with all the relevant information: spectra, NB images, HST counterparts, MUSELET results and redshift suggestions. With this information, users can decide on a redshift for an object as well as assigning each source a redshift-confidence level from 1 to 3. A confidence level of 1 denotes a tentative redshift and a confidence level of 3 denotes a redshift with a high confidence. For this study we only use LAEs with confidence levels of 2 and 3, indicating secure redshifts. In our sample -- A2744, A370, AS1063, M0416 -- there are 263 such LAEs. \\ 
\indent Magnifications of sources are assigned with the use of the parametric mass distribution models in \citetalias{richard2021atlas} (for A2744, A370 and M0416, for AS1063, the lens model comes from \citealt{beauchesne2023AS1063lensmodel}) and the LENSTOOL software \citep{kneib1996lenstooloriginal,jullo2007lenstool,jullokneib2009lenstool2,kneibnew2011lenstool}. These models are well constrained by the large number of multiple images with strong spectroscopic redshifts (levels 2 and 3 as described above) in these clusters. \citetalias{richard2021atlas} estimate a typical statistical error of $1\%$ of the mass profile of these clusters. The models in turn give the magnifications of the sources used in this study, which range from 0.8 (demagnified) up to 137. Most sources have magnifications between 1.5 and 25. While these lensing models are well understood and benefit from many multiple image systems as constraints, small systematic uncertainties related to the lens model used can still persist due to the particular choice of mass distribution (see, for example \citealt{Meneghetti2017HFF,Acebron2017lensing_systematics,Acebron2018strong_lensing_analysis,Furtak2021cluster_stellasmassfunc}). \\

\subsubsection{LAE Flux Determination}
\label{sect:laeflux}
The \lya flux for the LAEs in our sample was extracted using one of two different methods.\\
The main method employed in the LLAMA Sample is the line profile fitting procedure described in \citetalias{AC2022LLAMAS}. For the subsample used in this work (263 LAEs), all those behind A2744 and 20 behind the other three clusters (roughly half the subsample), we use a method involving SExtractor \citep{EB96SEx} following the procedure outlined in \citetalias{GdlV2019LAELF}. We give a brief description of both methods.\\
\indent The line profile fitting method utilises three steps: spectral fitting, NB image construction and spectral extraction. The first step involves fitting an asymmetric Gaussian function to the \lya peak with the EMCEE package from \citet{foreman2013emcee}, using 8 walkers and 10,000 iterations (in double peaked cases, both peaks are considered separately and their fluxes combined after the extraction is complete).  The peak position of the \lya line, flux, full width at half maximum (FWHM) and asymmetry of the source are all fitted. The second step takes the result for the peak position and creates a NB image around the LAE. The continuum around the LAE in redward and blueward bands of width $24\,\AA$ is subtracted from the \lya flux. The NB bandwidth is optimised such that the SNR of the \lya peak is maximised in an aperture of radius $0.7\arcsec$. Finally, utilising this new NB image of the LAE, a new extraction is performed from the MUSE datacube, ensuring that as much of the \lya emission from the galaxy and surrounding halo is extracted. The process is repeated twice more, each time using the latest NB image and extraction.\\
\indent The second method is employed for A2744 and in the three other clusters for the cases where the first method fails to fit the \lya flux. This happens in very low signal-to-noise cases or sources close to the edge of the datacube. This method uses SExtractor on NB images as described in \citetalias{GdlV2019LAELF}. The NB images in question are those in which each LAE is detected. Three sub-cubes are extracted from the main datacube, one encompasing the \lya emission and two either side of it (spectrally) each of width $20\,\AA$. These cubes are averaged to form a continuum image, which is subtracted pixel-by-pixel from an image formed by averaging the cube containing the \lya emission. SExtractor is then run on this new continuum-subtracted image and the {\tt FLUX\_AUTO} parameter is used to estimate the fluxes of the LAEs. To deal with faint sources or those with an extended, low surface brightness, SExtractor can progressively loosen the detection conditions (using the {\tt DETECT\_THRESH} and {\tt DETECT\_MINAREA} parameters) so that a flux can also be extracted for these sources. \\
Fig. 4 of \ttpaper shows the comparison between the two methods of extracting the \lya flux. In general, the two methods agree well. In some cases, the line profile fitting method estimates a larger flux due to its enhanced appreciation of the line complexity. \\
\indent In Fig.~\ref{fig:LAELBGzm1500dist} we show the properties of our sample of LAEs in red. In order to have a similar derivation of $\mathrm{M_{1500}}$ for all our LAEs, regardless of detection in HST photometry, values of $\mathrm{M_{1500}}$ are calculated from the filter closest to the $1500\,\AA$ rest frame wavelength (where available, and the $2\sigma$ upper limit of the continuum taken from this filter where not), and corrected for magnification. \lya luminosities are derived from the fluxes described above and also magnification-corrected. In terms of \lya luminosity we probe roughly between $39.5<\mathrm{log_{10}(L_{Ly\alpha}/erg\,s^{-1})}<42.5$, with decreased statistics near the faintest and brightest limits. This faint population can at present only be accessed with lensing clusters; the typical limit in blank fields lies around $\mathrm{log_{10}(L_{Ly\alpha})}\sim42$ (see for example, \citet{herenz2019MUSELAELF}), down to $\mathrm{log_{10}(L_{Ly\alpha})}\sim41.5$ in the MUSE Ultra Deep Field \citep{drake2017MUSELAELF}.

\subsection{HFF data}
\label{sect:HFF}
The HST observations of the clusters used in this work are from the Hubble Frontier Fields program (ID: 13495, P.I: J. Lotz). In particular, we use the photometric catalogues of the HFF-DeepSpace Program \citep{shipley2018hff}. As part of this program, the authors collected homogeneous photometry across the HFF, including deep $K_{S}$-band imaging ($2.2$\mm) from the HAWK-I on the Very Large Telescope (VLT) and Keck-I MOSFIRE instruments \cite{Brammer2016KeckHFF}. Additionally included are all available data from the two Spitzer/IRAC channels at 3.6\mm and 4.5\mm. The two Spitzer/IRAC channels at 5.8\mm and 8\mm were judged to be too noisy and excluded from the SED fitting process (see Section~\ref{sect:LBG_selection}). The details of the filters used, and their depths, for each cluster can be found in Table~\ref{table:filter_depths}. The bright cluster galaxies (BCGs) and intra-cluster light (ICL) are subtracted by \cite{shipley2018hff} for improved photometry of background sources in these very crowded fields. The detection image for each of these clusters is made up of a combination of the F814W, F105W, F125W, F140W and F160W bands (PSF-matched to the F814W band), after the modelling out of the BCGs. \\
\indent The area that MUSE observed for each of these clusters is fully contained within the HST area, so we cut the HST area we consider to that of the MUSE data. The resulting effective (lens-corrected) co-volume, derived from LENSTOOL source-plane projections, is $22910\,\mathrm{Mpc^{3}}$ over the redshift range $2.9<z<6.7$.

\begin{table}
    \caption{$5\sigma$ limiting magnitudes of each filter for the four clusters.}
    $$
    \begin{array}{ p{0.15\linewidth} p{0.1\linewidth} p{0.1\linewidth} p{0.12\linewidth} p{0.1\linewidth} }
        \hline
        \hline
        \noalign{\smallskip}
        \textbf{Filter} & \textbf{A2744} & \textbf{A370} & \textbf{AS1063} & \textbf{M0416} \\
        \hline
        \noalign{\smallskip}
        F225W & -- & -- & 24.0  &  24.4 \\
        \hline
        \noalign{\smallskip}
        F275W & 26.0 & 25.7 & 25.9 & 25.7 \\
        \hline
        \noalign{\smallskip}
        F336W & 26.7 & 26.2 & 26.3 & 26.2\\
        \hline
        \noalign{\smallskip}
        F390W & -- & -- & 25.4 & 25.7\\
        \hline
        \noalign{\smallskip}
        F435W & 27.5 & 27.2 & 27.3 & 27.5\\
        \hline
        \noalign{\smallskip}
        F475W & -- & 26.6 & 25.8 & 26.2\\
        \hline
        \noalign{\smallskip}
        F606W & 27.7 & 27.3 & 27.5 & 27.7\\
        \hline
        \noalign{\smallskip}
        F625W & -- & 25.6 & 25.5 & 25.9\\
        \hline
        \noalign{\smallskip}
        F775W & -- & -- & 25.5 & 25.6\\
        \hline
        \noalign{\smallskip}
        F814W & 27.8 & 27.5 & 27.7 & 27.9\\
        \hline
        \noalign{\smallskip}
        F850LP & -- & -- & 25.2 & 25.2\\
        \hline
        \noalign{\smallskip}
        F105W & 27.9 & 27.2 & 27.6 & 27.7\\
        \hline
        \noalign{\smallskip}
        F110W & -- & 26.6 & 26.4 & 26.3\\
        \hline
        \noalign{\smallskip}
        F125W & 27.4 & 27.0 & 27.4 & 27.6\\
        \hline
        \noalign{\smallskip}
        F140W & 27.4 & 27.1 & 27.3 & 27.7\\
        \hline
        \noalign{\smallskip}
        F160W & 26.8 & 26.9 & 26.9 & 27.4\\
        \hline
        \noalign{\smallskip}
        K\textsubscript{S} & 25.1 & 24.9 & 25.1 & 25.1\\
        \hline
        \noalign{\smallskip}
        IRAC1 & 24.1 & 23.6 & 23.6 & 24.7\\ 
        \hline
        \noalign{\smallskip}
        IRAC2 & 24.3 & 23.7 & 24.2 & 24.2\\
        \hline
        \hline
    \end{array}
    $$
    \begin{tablenotes}
      \small
    \item \textbf{Notes.} Values have been calculated from the properties of the sample and adopted for the SED fitting described in Section.~\ref{sect:LBG_selection}. Dashes indicate missing filters for a particular cluster.
    \end{tablenotes}
    \label{table:filter_depths}
\end{table}

\subsection{LBG selection}
\label{sect:LBG_selection}
\begin{figure}
    \centering
    \includegraphics[width=0.5\textwidth]{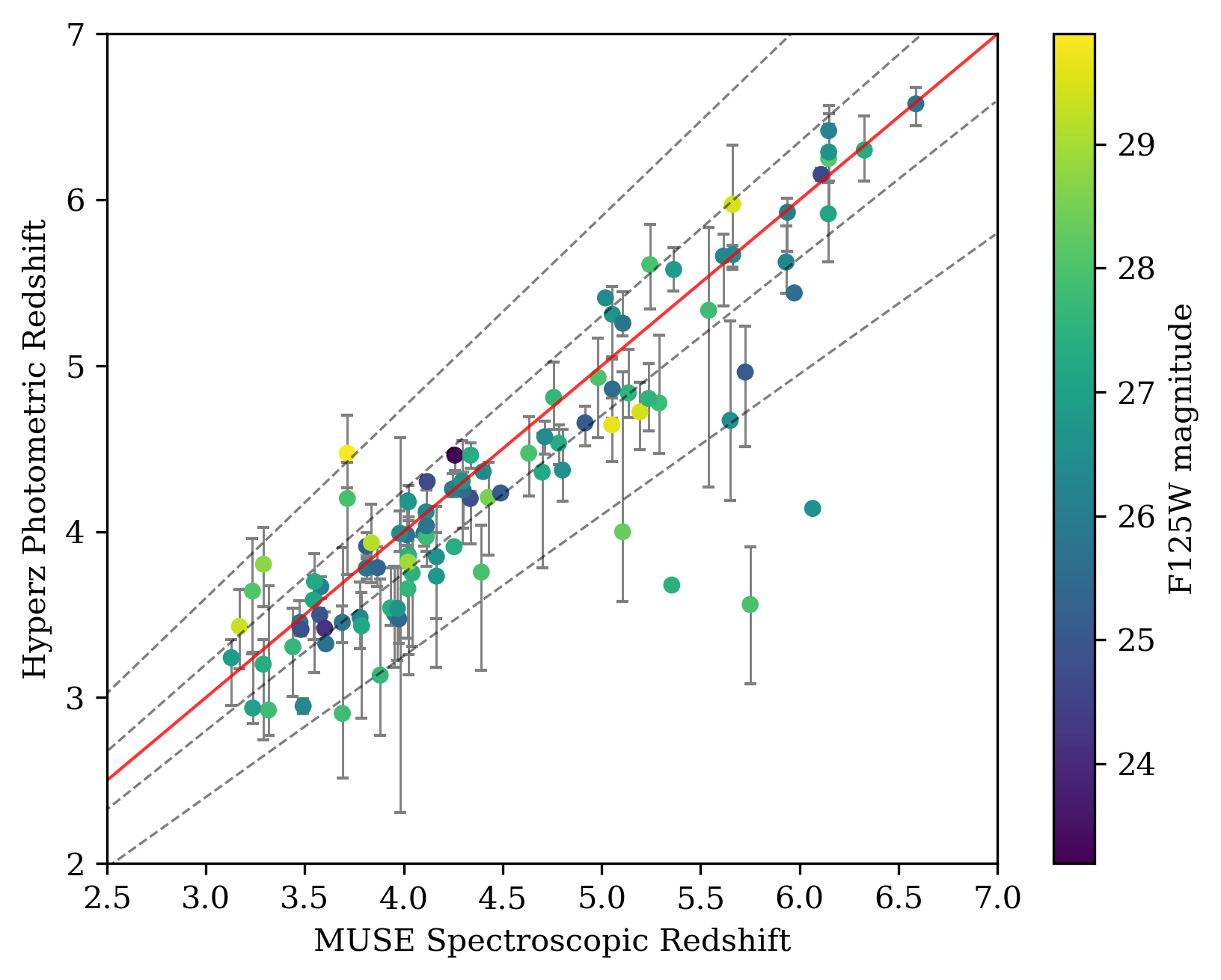}
    \caption{Photometric vs Spectroscopic redshift comparison of objects selected as LAEs and LBGs. The red line indicates the one-to-one relation and the dashed black lines indicate the typical error bounds of $|z_{spec}-z_{phot}|<0.15(1+z_{spec})$ and $|z_{spec}-z_{phot}|<0.05(1+z_{spec})$. Almost all of the photometric redshifts agree with their spectroscopic equivalents within the outer of these two bounds. Those that do not tend to be fainter objects, with magnitudes of 28 and above.}
    \label{fig:z_compare}
\end{figure}

We calculate photometric redshifts and probability distributions (hereafter $P(z)$) in order to perform our LBG selection. We use the package \hypz \citep{MB00hyperz} to estimate redshifts and probability distributions. This package uses a standard $\chi^2$ minimisation technique to fit template galaxy Spectral Energy Distributions (SEDs) to photometric data points. It has been optimised for the redshift determination of high-redshift galaxies so is ideal for our purpose. We use a suite of template spectra to fit the photometric data: the four template spectra from \citet{CWW1980colors}, two Starburst99 models with nebular emission \citep{leitherer1999starburst99}: a single burst model and a constant star-formation rate model (where each spans five metallicities and 37 stellar population ages), and seven models adapted from \citet{BandC03SED}. Included in these seven are a star formation burst model, five exponentially decaying models with timescales between 1 and $30\,\mathrm{Gyr}$ and a constant star formation model. The redshift range we use in \hypz is 0 to 8 with a step in redshift of 0.03. The Calzetti extinction law \citep{Calzetti00dust} is used to account for internal extinction with values of $A_v$ allowed to vary between 0 and 1.5 magnitudes. No priors on galaxy luminosity were introduced during this process, avoiding any bias introduced by lensing magnification.\\ 
\indent For a galaxy to be included in the LBG sample for this study we demand that the best solution for the photometric redshift lies in the range $2.9<z<6.7$. We also accept any candidates not in this range which have $1\sigma$ errors overlapping with it. In practice, this accounts for just 14 objects. To make sure the detection is real, we also demand a $5\sigma$ detection in at least one HST filter. \hypz provides a redshift probability distribution, $P(z)$, which can have two peaks, in which case there are two redshift solutions for that particular galaxy. We accept galaxies into the sample if only one of the peaks is in the required range, as long as $60\,\%$ of the integrated $P(z)$ also lies in this range. We can compare the photometric redshift results for those objects also selected as LAEs to the spectroscopic redshift determined from the peak of the \lya emission. Fig.~\ref{fig:z_compare} shows this comparison. The dashed lines indicate regions constrained by $|z_{spec}-z_{phot}|<0.15(1+z_{spec})$ and $|z_{spec}-z_{phot}|<0.05(1+z_{spec})$. We encode the apparent magnitude in the F125W filter in the colourbar to demonstrate that the instances in which \hypz performs poorly (i.e. instances outside of the outer region specified above) tend to be fainter objects with magnitudes $\gtrsim28$.\\
\indent Finally, we manually inspect our LBG sample using the HST images, photometry, SED fitting from \hypz and the LENSTOOL package with the lens models described in Section~\ref{sect:laeselection}. This procedure is designed to identify multiple images in the LBG sample caused by the gravitational lensing, as well as to remove any obvious spurious detection; regions of noise or contamination erroneously selected as LBGs. Multiple images are identified using a combination of LENSTOOL predictions, redshift determinations and SEDs from \hypz, and visual inspection of the HST images. LENSTOOL predictions, redshift estimates and colours of objects have to match for those objects to be designated a multiple image system. The colours used are $\mathrm{F814W-F606W}$, $\mathrm{F105W-F814W}$, $\mathrm{F140W-F105W}$, all of which have to match to within $2\sigma$ for the objects in question to be designated multiple images. From each system, a representative LBG image is chosen. This image is the least contaminated, or, in the case of a match with an LAE multiple image system, the image chosen to represent the LBG multiple system matches the selected LAE image (see Section~\ref{sect:matching}). \\

\subsection{Matching populations}

\label{sect:matching}
Having blindly selected our LAE and LBG samples, we compare them, using a matching radius of $0\farcs5$, to see which objects are selected as both. The results are shown in Table~\ref{table:data_details}. \\
We select a population labelled LAE+continuum, which denotes objects that are selected as LAEs, for which we see a continuum in the HST images, but where this continuum fails to meet the selection criteria for our LBG sample. Mostly, these galaxies fail on the SNR criteria outlined in the previous subsection, indicating a faint continuum or a noisy area in the HST images. Nevertheless, we keep these objects as, thanks to the \lya emission, we know that there is a high-redshift galaxy at these positions. In the previous work of this nature solely on A2744, \citetalias{GdlV2020LAEfrac} simply included these objects in the LAE+LBG sample, relaxing the signal-to-noise criterion for these objects, seeing as the presence of an object (as well as its redshift) was known thanks to the \lya emission. Here however, we keep the distinction between these continuum detections with \lya emission and our LBG sample which fulfil all the criteria laid out in the previous subsection. Details on the inclusion of this sample into the analysis of the LAE fraction, \xlae, are given in Section~\ref{sect:LAEfraction}.\\
\indent For the LAEs, a `best' image is chosen, a process described fully in \ttpaper. This allows us to chose representative images that are minimally impacted by neighbours, have high signal-to-noise and moderate magnification. When selecting LBGs, we keep the image corresponding to the best LAE image selected by \ttpaper where possible. However, in some cases we chose another image, because that particular LAE image is selected as LBG rather than continuum only. We ensure that the image chosen is always of similar quality. A modification of this nature is rare and we impose it in less than 10 systems across the whole LAE sample. The original \lya flux and magnification of the source, as given in the LLAMAS catalogues, is used for our analysis on the properties of these objects, however their designation as LAE+continuum or LAE+LBG can change based on this. \\

\subsection{Completeness Determination}
\label{sect:LAEcompleteness}
As covered in Section~\ref{sect:intro}, the completeness of the populations considered in such a study is very important. This correction to the number of sources detected aims to reflect the number of sources that are missed in the detection process and thus the number of sources actually present in the field of view. \\
The completeness methods used for the LAEs in this work are described in \ttpaper and summarised here. Following the procedure first laid out in \citetalias{GdlV2019LAELF}, sources are treated individually in this computation. Each source's brightness distribution profile, both in the spatial and spectral dimensions, is modelled and randomly injected 500 times into the NB layer of the original MUSE datacube where its \lya emission reaches a maximum. This process is performed in the image plane, in order to recreate as closely as possible the actual process of detection with MUSELET. The completeness of the source is the number of times (out of 500) it is successfully detected and extracted. \\
\indent Since it is the local noise that likely decides whether or not such an injected source is detected, to account for variations in the local noise, \ttpaper change the size of the NB image used to re-detect the simulated sources from $30\arcsec\times30\arcsec$ to $80\arcsec\times80\arcsec$. This was found to improve the extraction of the source when that source has close neighbours. The mean completeness value found for our sample is 0.72 and the standard deviation is 0.34. \\
Finally, each source's contribution to the LAE fraction is corrected by a value $1/\mathrm{C_i}$ where $\mathrm{C_i}$ is the completeness value of that source. In effect: a source with a completeness value of 0.5 contributes 2 LAEs to the calculation of the LAE fraction. 

\section{Results}
\label{sect:results}

\begin{figure}
    \centering
    \includegraphics[width=0.5\textwidth]{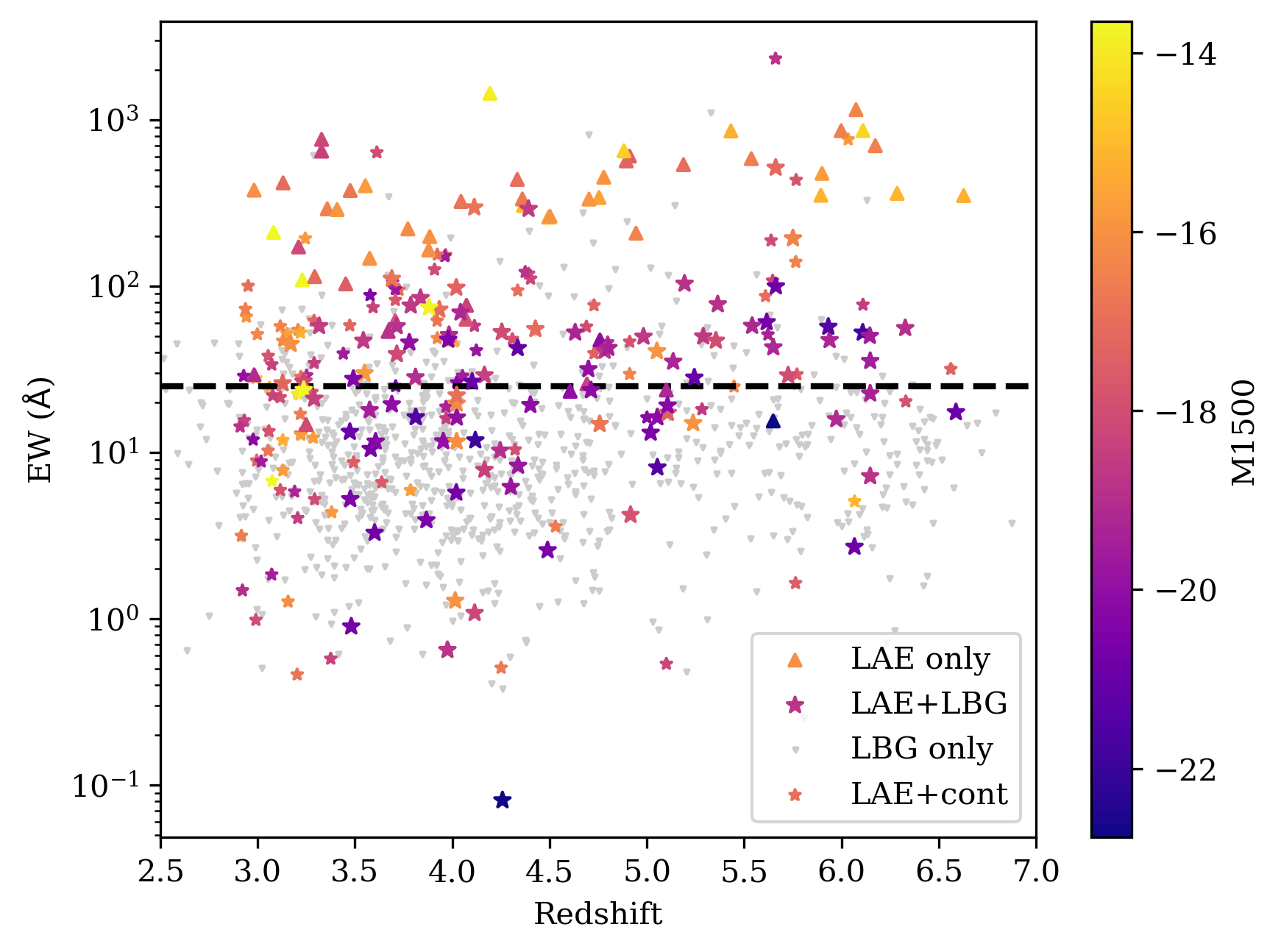}
    \caption{\lya equivalent width (\ewlya) distribution for all our samples. Objects selected as LAE and LBG are denoted by larger stars, LAE+continuum by smaller stars, LAE only by upturned triangles and LBG only by downturned triangles. \ewlya values for the latter two populations are calculated using the upper limits of the continuum and \lya flux respectively. The objects are colour-coded by $\mathrm{M_{1500}}$. The horizontal dashed line demarcates the $25\,\AA$ level above which LAEs are included in the calculation of the LAE fraction. Error bars are omitted for clarity, but shown on Fig.~\ref{fig:EWvszM1500} and \ref{fig:EWvsbeta}.}
    \label{fig:EWvsz}
\end{figure}

\begin{figure}
    \centering
    \includegraphics[width=0.5\textwidth]{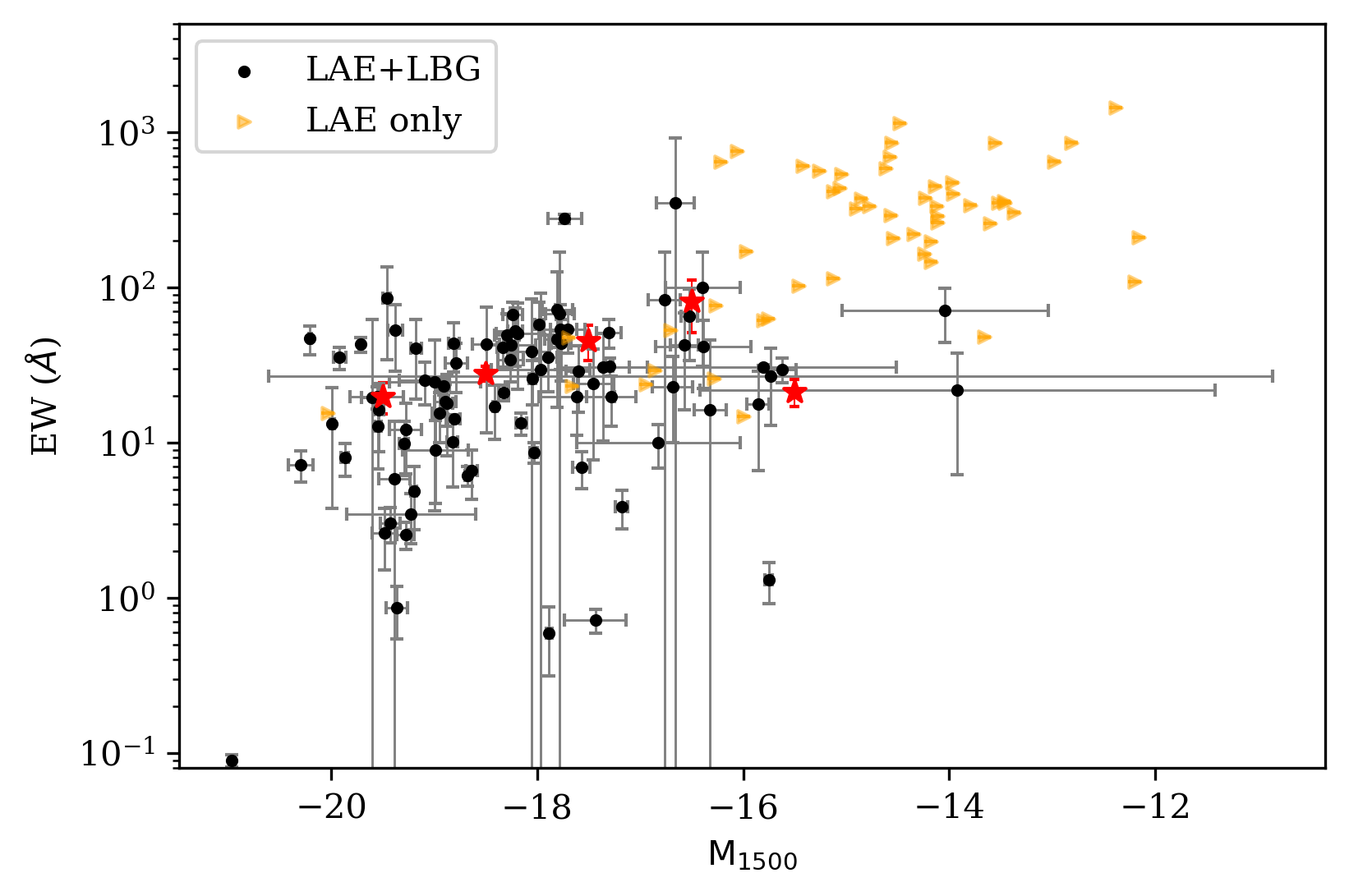}
    \caption{The \ewlya distribution over the range of UV absolute magnitude probed for the LAE+LBG sample. The red stars indicate the average in five equal--size bins between $\mathrm{M_{1500}}=-20$ and $\mathrm{M_{1500}}=-15$, the region which is well populated. \cite{deBarros2017LAEfraction} (at $z=6 $) provide a similar plot for a brighter sample with a similar rising average \ewlya. Our plot extends several magnitudes fainter and we see that this trend continues down to at least $\mathrm{M_{1500}}=-16$. An idea of the region fainter than $\mathrm{M_{1500}}=-15$ can be gained by including the LAE only sample (orange triangles), the vast majority of which are very faint in UV magnitude, high-\ewlya objects. (For these objects, EW error bars, coming only from the error on the \lya flux, are smaller than the size of the points. Continuum values used are $2\sigma$ upper limits estimated from the filter that would see the emission at $1500\,\AA$ rest frame).}
    \label{fig:EWvszM1500}
\end{figure}

\begin{figure}
    \centering
    \includegraphics[width=0.5\textwidth]{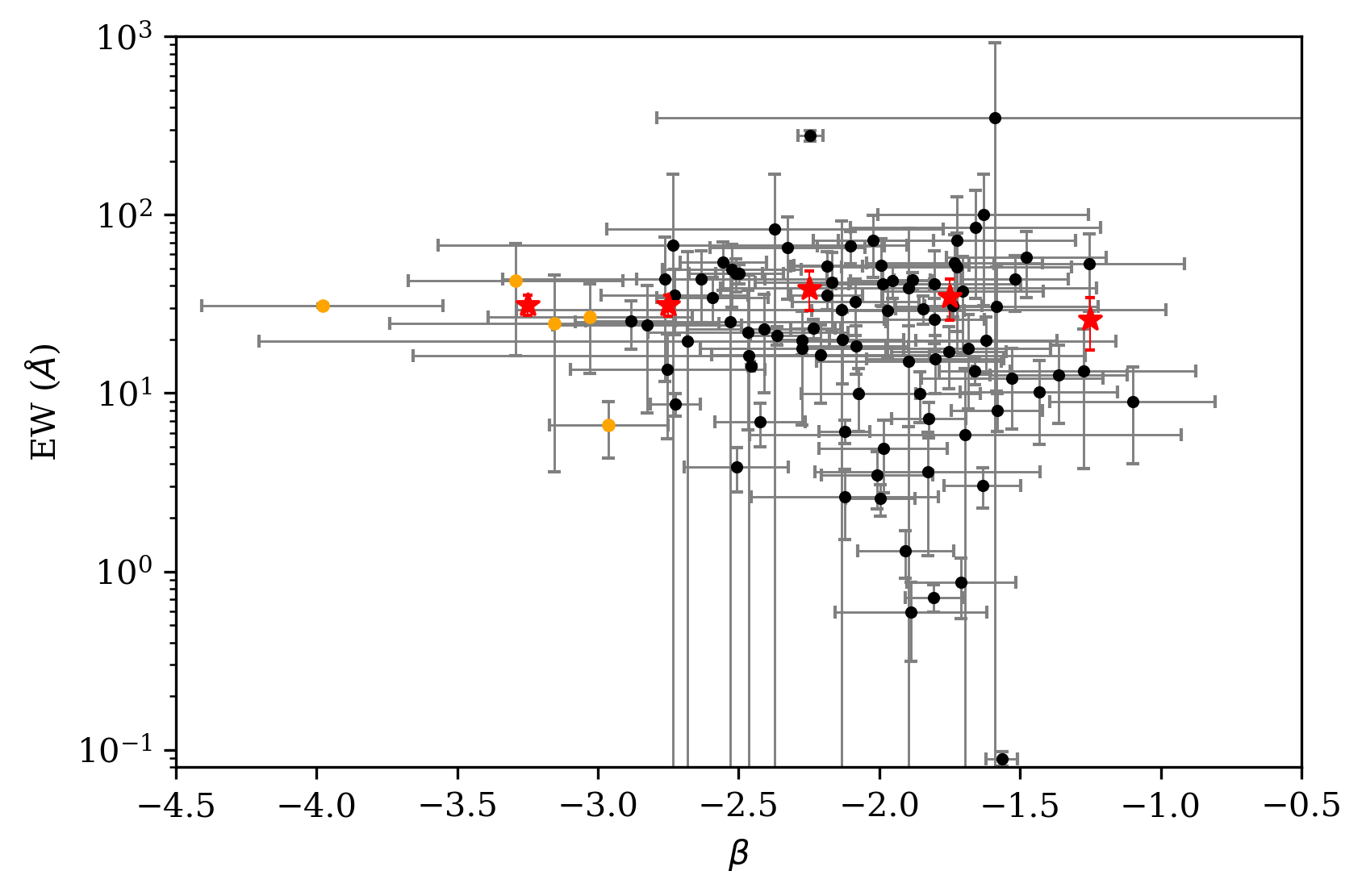}
    \caption{UV slope plotted against \ewlya for the LAE+LBG sample. Orange dots represent UV slopes which, upon inspection, we find to be dubious: photometry which locally does not represent the slope well or with large errors. Red stars represent average values in equally sized bins between $\beta=-3.5$ and $\beta=-1.0$. The change of \ewlya across this range is not as significant as the change across the range of UV magnitudes probed (see Fig.~\ref{fig:EWvszM1500}), however we see a lightly rising trend in \ewlya with bluer slopes down to $\beta=-2.5$.}
    \label{fig:EWvsbeta}
\end{figure}

\subsection{UV properties of the Populations}
\label{sect:UVprop}
\begin{figure}
    \centering
    \includegraphics[width=0.5\textwidth]{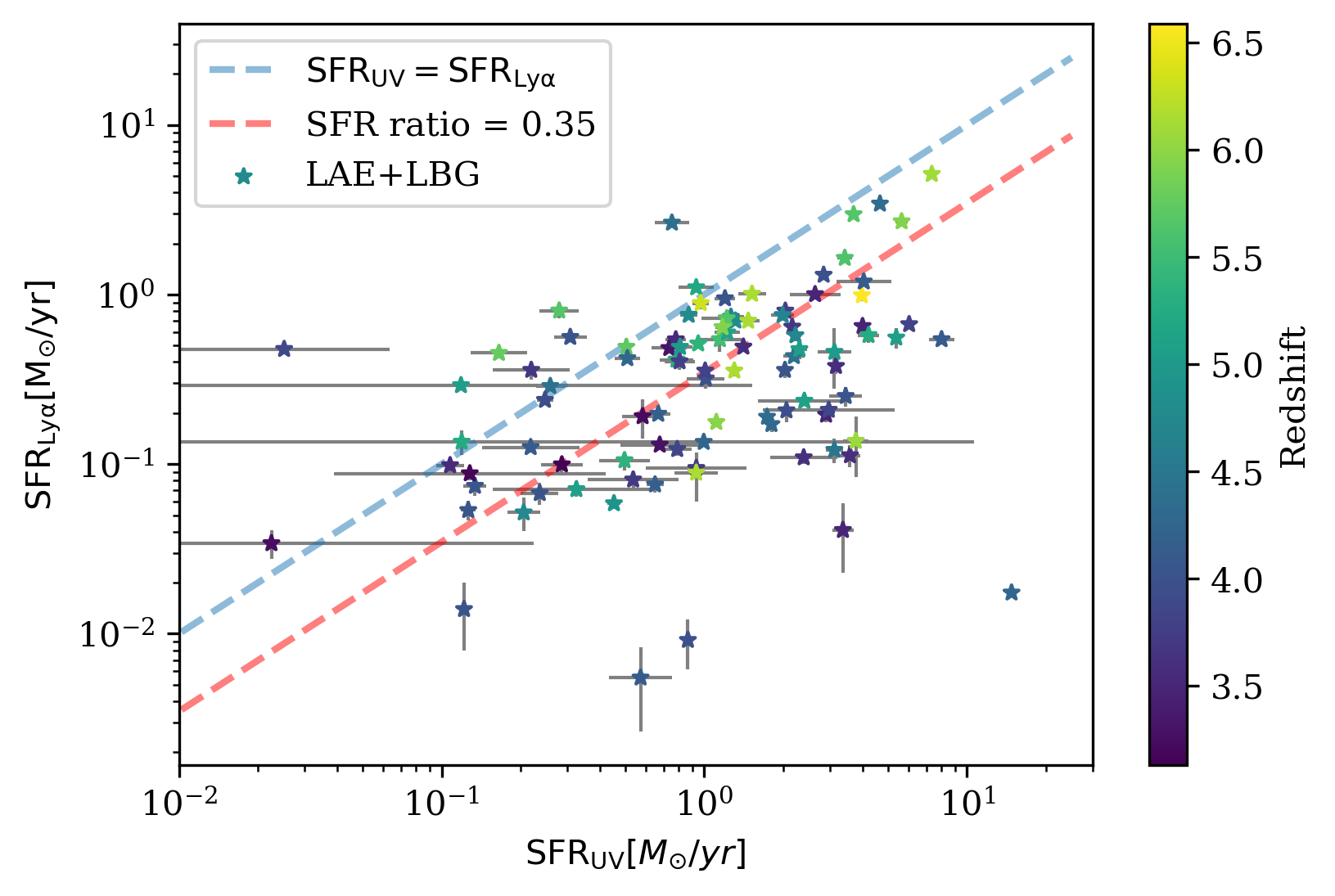}
    \caption{Comparison of the dervied SFRs for our LAE+LBG sample, colour-coded by redshift. The errors on the \sfrlya are smaller than the size of the points in most cases. The blue dashed line denotes a one to one SFR ratio, i.e. \sfruv = \sfrlya. Galaxies along this line also have equal escape fractions: \fuv = \flya. Most of our sample lies under this line and the red dashed line denotes the median SFR ratio of our sample, 0.35. For fainter objects with \sfruv$<1\,M_{\odot}/\mathrm{yr}$, a greater fraction lie above the line of equal SFR, a trend previously observed in \citet{ando2006stronglyadef,schaerer2011LAEfrac} and \citetalias{GdlV2020LAEfrac}. }
    \label{fig:SFR}
\end{figure}

\begin{figure}
    \centering
    \includegraphics[width=0.5\textwidth]{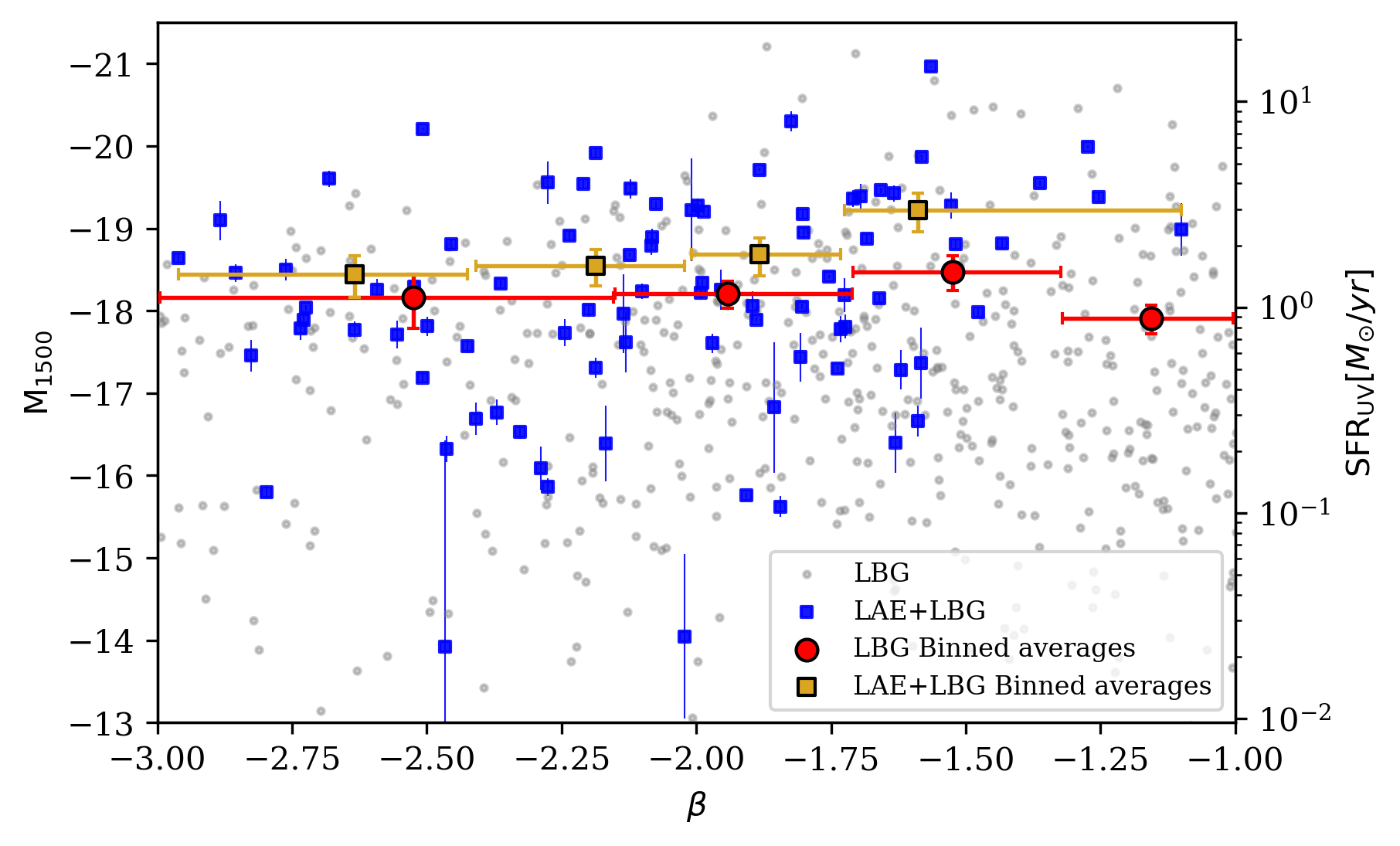}
    \caption{
    $\mathrm{M_{1500}}$ and \sfruv vs. UV slope. The sample selected as LBG only is shown in grey and the LAE+LBG sample is colour-coded by redshift. Binned averages in \sfruv for the LBG and LAE+LBG samples are shown by red circles and golden squares respectively. Each bin contains an equal \textit{number} of the respective samples and the horizontal error bars show the width (in terms of $\beta$) of each bin. Errors on the LBG only objects are omitted for readability and we note that there are LBG only objects (and 4 LAE+LBG objects; see Fig.~\ref{fig:EWvsbeta}) that extend beyond the $\beta$ limits of this graph, however these are objects where the calculation of the UV slope is suspicious. There are LBG only objects that extend beyond the $\mathrm{M_{1500}}$ limits of the graph but these are taken into account when computing the binned averages.}
    \label{fig:SFRvsbeta}
\end{figure}

In order to ascertain the similarities and differences between our populations of high redshift galaxies, it is useful to look at the physical properties we derive from the HST photometry and their relation to \lya emission. This can also help to disseminate how these properties tie in with the LAE fraction (see Section~\ref{sect:laeflux}). Additionally, we can appraise any differences from the established trends for these high redshift galaxies that may appear in our sample of faint, lensed galaxies.\\
\indent To evaluate this relation to \lya emission, the \lya equivalent width, \ewlya, is an important property to derive. We can then compare this to UV-derived properties. In order to calculate \ewlya values for our sample, first we derive UV slopes by fitting the photometric points starting above the location of the \lya emission (irrespective of whether or not we detect it for a given galaxy) and including all the filters up to $2600\,\AA$ in the rest-frame, adapting the approach used in \citet{castellano2012Beta,schenker2014lyafracfeasibility,bouwens2015UVLF}. We fit a power law, $F_{\lambda}\propto\lambda^{\beta}$, where $\beta$ is the UV slope, to the photometry. We choose this method as it does not rely on SED fitting with a set of templates which have specific allowed values of $\beta$.\\
\indent Subsequently, by using this photometrically fitted UV slope for each object, we can ascertain the continuum flux level beneath the \lya emission. Hence we calculate the \lya equivalent width (\ewlya) by dividing the \lya flux by the continuum flux level, corrected to its rest-frame level. In this process we take into account the error on the UV slope resulting from the fitting process, as well as the error associated with the \lya flux (see Section~\ref{sect:LAE}).\\
\indent For the objects with no associated continuum, the upper limits of the continuum are taken from the filter closest to the $1500\,\AA$ rest-frame emission. For the objects with no associated \lya emission, the detection limits of the \lya emission are used.\\
\indent The EW distribution with redshift for all our sample is shown in Fig.~\ref{fig:EWvsz}. The dashed black line denotes the typical EW inclusion limit for objects to the calculation of \xlae (see Section~\ref{sect:LAEfraction}). The percentage of objects above this limit is $44\%$ and $50\%$ for LAE+LBG objects and LAE+continuum objects respectively. This is to be expected as the objects selected as continuum but not LBG are in general fainter, giving rise to higher values of \ewlya.\\ 
\indent In Fig.~\ref{fig:EWvszM1500} we compare \ewlya to UV magnitude for our LAE+LBG sample. UV absolute magnitude, $\mathrm{M_{1500}}$, is calculated from the filter closest to the $1500\,\AA$ rest-frame emission. We see a rise in \ewlya towards fainter $\mathrm{M_{1500}}$, in agreement with results reported in \citet{stark2010keckLAEfrac,deBarros2017LAEfraction,kusakabe2020}. We note that above $M_{1500}=-16$ this graph is populated by high EW, continuum-undetected LAEs (shown by orange triangles). We do not include these objects in the binned averages as these objects are not LBGs and the continuum values are estimated upper limits, however these objects indicate that this trend in EW likely increases to even fainter magnitudes than is populated by our current LBG selection. As the spatial extent of \lya emission correlates with the size of the galaxy and hence $\mathrm{M_{1500}}$ (\citealt{wisotzki2016extendedlyahaloes,leclercq2017lyahaloes}; \citetalias{AC2022LLAMAS}), the trend in Fig.~\ref{fig:EWvszM1500} is often seen as a natural result. \\
\indent We also compare \ewlya to the UV slope (Fig.~\ref{fig:EWvsbeta}). In general, high-EW LAEs are found to have bluer slopes \citep{stark2010keckLAEfrac,schenker2014lyafracfeasibility,deBarros2017LAEfraction}. We report a slight increase ($<1\sigma$) in average EW in bins between $\beta=-1.0$ and $\beta=-3.0$. However, similarly to \citet{deBarros2017LAEfraction} we find \ewlya values as high as $\sim100\,\AA$ across the sample. \\
\indent  We derive star-formation rates from \lya (\sfrlya) and from the UV continuum (\sfruv), based on the relations given in \citet{kennicutt1998Schmidtlaw} and the factor of 8.7 between $\mathrm{L(H\alpha)}$ and $\mathrm{L(Ly\alpha)}$, assuming a Salpeter IMF \citep{salpeter1955IMF} and constant star formation.
\sfrlya is a good lower limit on the intrinsic SFR as \lya flux can be lost due to dust attenuation or a more or less opaque IGM \citep{zheng2010lya_radiativetransfer,gronke2021lyatransmission}. However due to the use of an IFU we have no impact on \lya flux from slit losses.\\
\indent We compare the two measures of SFR for our LAE+LBG sample in Fig.~\ref{fig:SFR}, plotting the line of one to one ratio and the actual median ratio (\sfrlya/\sfruv) found in our sample. This median ratio (0.35), is well below the one to one ratio, indicating that in most cases the escape fraction of UV photons (the fraction of photons at $1500\,\AA$ rest frame that escape the galaxy), \fuv, exceeds that of \lya photons (the fraction of \lya photons that escape the galaxy), \flya. This is less apparent in UV-fainter objects, where there is a greater fraction above the \fuv = \flya line. This result is in line with previous findings in \citet{ando2006stronglyadef,schaerer2011LAEfrac} and \citetalias{GdlV2020LAEfrac}. The explanation offered previously by \citet{ando2006stronglyadef} and \citetalias{GdlV2020LAEfrac} relates to the likely difference between the UV-bright and UV-faint galaxies. If the UV-bright galaxies have had more time to evolve than UV-fainter objects, it is likely that they are chemically more complex and have a larger amount of dust. This would decrease the escape of \lya photons and hence result in a smaller measured \sfrlya. We return to this point having compared the LAE fraction for the UV-bright and UV-faint halves of our sample in Section~\ref{sect:LAEfraction}.\\
\indent Subsequently, we compare $\mathrm{M_{1500}}$ and \sfruv to the UV slopes of both samples (LBG and LAE+LBG) between $-3.0<\beta<-1.0$: Fig.~\ref{fig:SFRvsbeta}. In this graph, we calculate binned averages as in Figs.~\ref{fig:EWvszM1500} and \ref{fig:EWvsbeta}, however for this plot the averages are computed for bins of equal population. For the LBG only sample, there is a slight increase in \sfruv with steeper UV slopes, this is to be expected as steeper values of $\beta$ generally indicate populations that are more intensely star forming.\\
From the binned averages we can see that for the population also selected as LAEs, the galaxies are distributed towards higher values of \sfruv and steeper UV slopes. This result supports previous findings (in various redshift ranges, as well as in simulations) that LAEs are in general dust-poor and are more intensely star forming than objects only selected as LBGs \citep{verhamme2008lyadustsim,hayes2011redshiftevofdust,hayes2013lyadust,sobral2019predictinglyaesc,santos2020lyaUVproperties}.\\
\indent We extend this genre of analysis to our large sample of intrinsically faint galaxies, finding the same trends; LBGs also selected as LAEs have on average higher \sfruv and steeper UV slopes. High-\ewlya objects are on average fainter in absolute UV magnitude with steeper UV slopes, and those fainter in UV absolute magnitude tend to have \flya approaching \fuv or in some cases above, while this is not seen for UV-brighter LAEs. All these trends are seen with significant scatter, to be expected given the intrinsically faint nature of the sample, the limits of the photometry (albeit the deepest available) and those objects that have a complex star formation history and physical structure, for which our assumption of constant star formation is less valid. Combined, our results reinforce the picture of high-redshift LAEs as UV-faint, intensely star forming, dust-poor galaxies and show that these trends hold for galaxies down to $\mathrm{M_{1500}\sim-12}$.

\subsection{The Interrelation between the LAE and LBG populations}
\label{sect:interrelation}

\begin{figure*}
    \centering
    \includegraphics[width=0.8\textwidth]{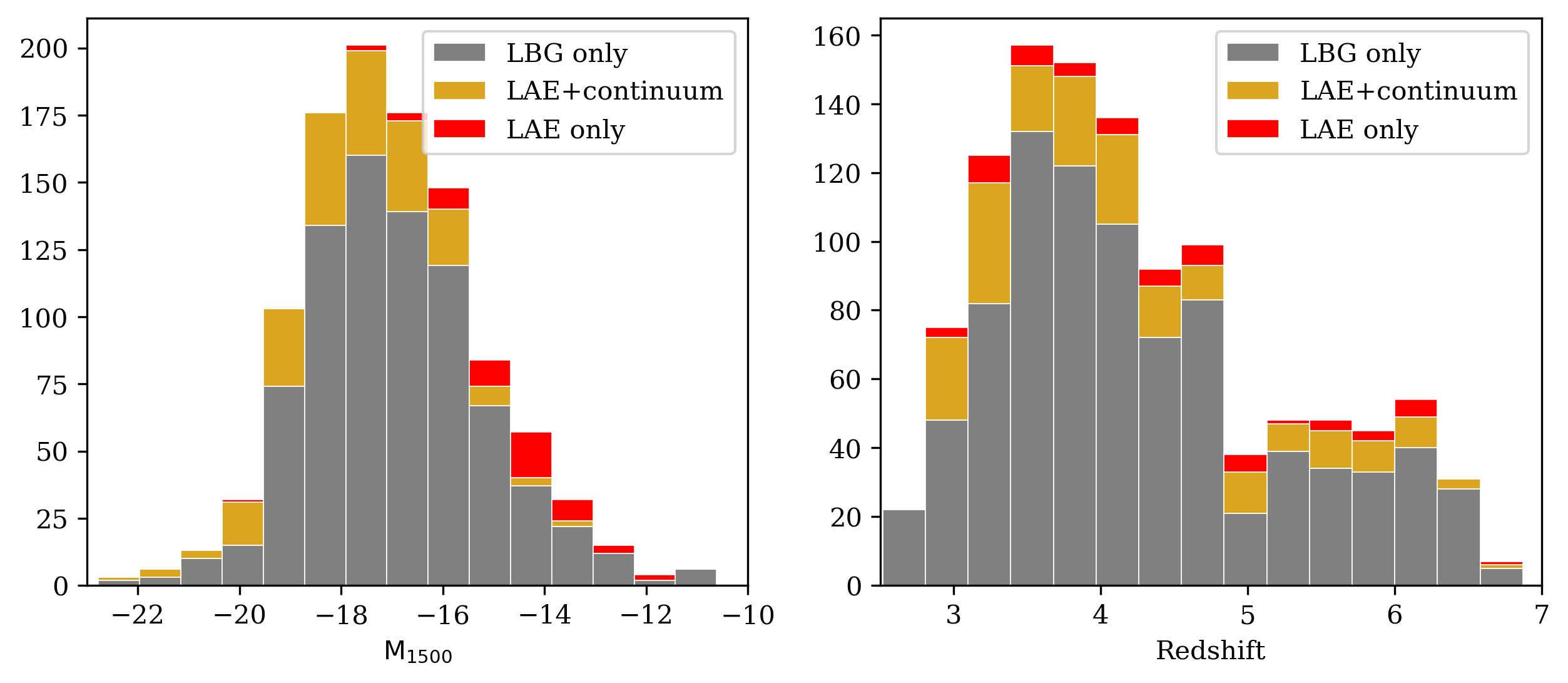}
    \caption{$\mathrm{M_{1500}}$ distribution of the populations \textit{(left)} and redshift distribution of the populations \textit{(right)}. The sample here labelled as `LAE+continuum' includes all LAEs matched with continuum objects and LBG--selected galaxies.}
    \label{fig:M1500_z_distribution}
\end{figure*}

We define four samples: objects identied as LBG only (henceforth LBG only), objects identified as LAE only (henceforth LAE only), objects identified as both LAE and LBG (LAE + LBG) and objects identified as LAE which have a continuum detected in the photometry from HST which itself does not meet our criteria for the LBG sample (LAE+continuum). These samples have 872, 55, 100 and 108 members respectively. \\
\indent The sample is dominated by A2744 with $46\%$ of the LAEs and $35\%$ of the LBGs; this is a large field of view with long MUSE exposures (see \citetalias{richard2021atlas}, \citetalias{AC2022LLAMAS} and \ttpaper), however, as many fields of view as possible are valuable additions to the sample. The redshift distribution and the $M_{1500}$ distribution of the sample are displayed in Fig.~\ref{fig:M1500_z_distribution}. We find the same result as \citetalias{GdlV2020LAEfrac} when considering the relative importance of LAEs which are totally undetected in the continuum. Their increasing presence is clear when looking at the faint region of the $\mathrm{M_{1500}}$ distribution, even when using the deepest HST photometry. This is crucial to note for any study wishing to catalogue the contribution to reionisation of all star-forming objects at high-redshift. This effect is naturally expected to strengthen with decreasing depth of photometry.

\subsubsection{LAE fraction}
\label{sect:LAEfraction}

\begin{figure}
    \centering
    \includegraphics[width=0.5\textwidth]{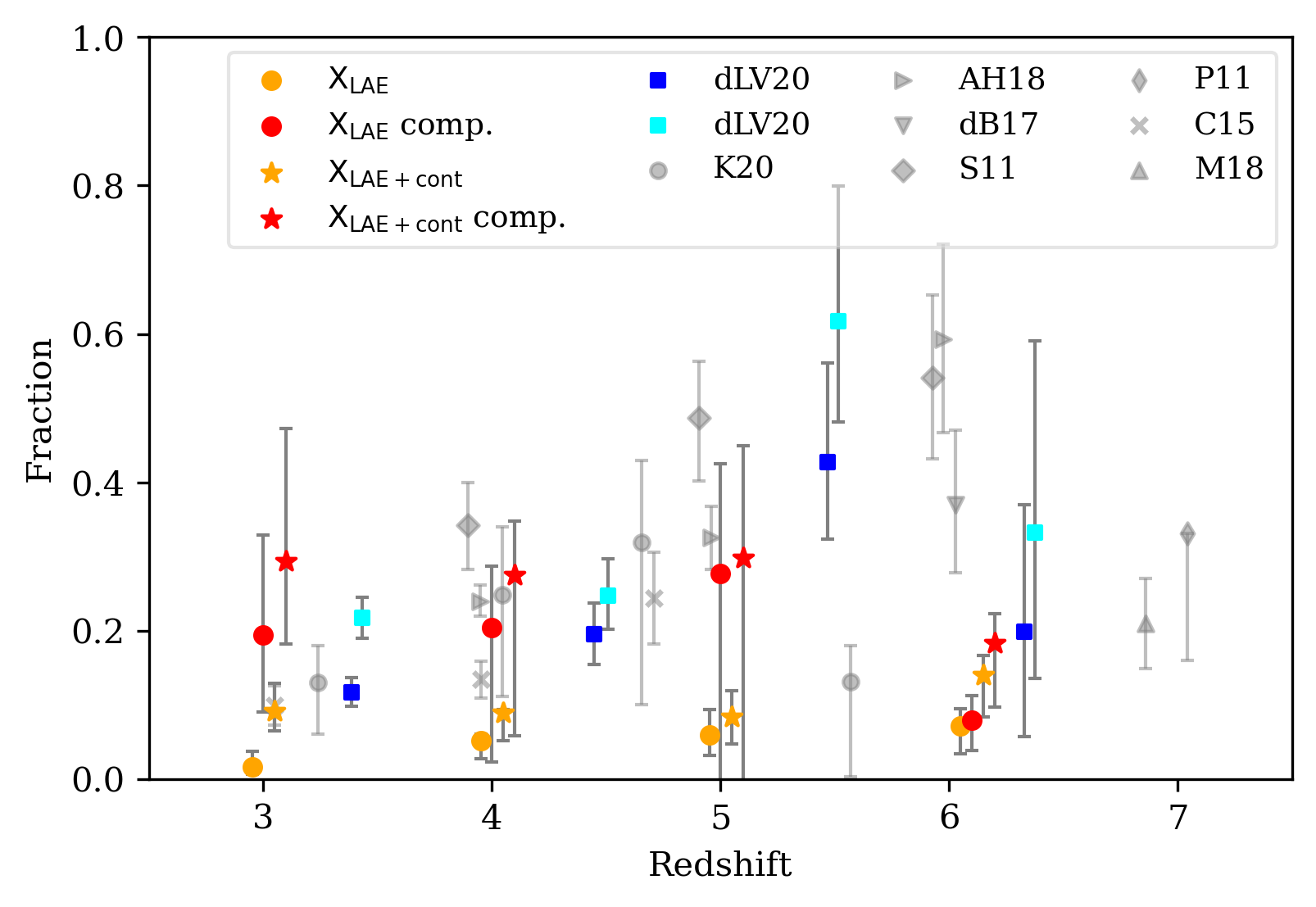}
    \caption{The LAE fraction, \xlae, over a redshift range of 2.9 to 6.7, using the typical literature limits of \ewlya$>25\,\AA$ and $\mathrm{M_{1500}}>-20.25$ respectively. The coloured circles represent \xlae counting just LAE+LBG objects and the coloured stars represent the same calculation including the LAE+continuum sample (see Section~\ref{sect:matching}. Red stars indicate \xlae inclusive of the completeness correction on the number of LAEs (see Section~\ref{sect:LAEcompleteness}). Blue and cyan squares represent the results from \citetalias{GdlV2020LAEfrac} with cyan representing those with the completeness correction. Shorthand is used in the legend for the literature results but we give the full list here (in the same order as the legend): \cite{kusakabe2020,arrabal2018LAEfrac,deBarros2017LAEfraction,stark2011LAEfrac,pentericci2011laefrac/z=7LBG,cassata2015vimosLAEfrac,mason2018LAEfrac}. A small redshift offset is artificially applied to some points for clarity.}
    \label{fig:LAEfrac20.25}
\end{figure}

\begin{figure}
    \centering
    \includegraphics[width=0.5\textwidth]{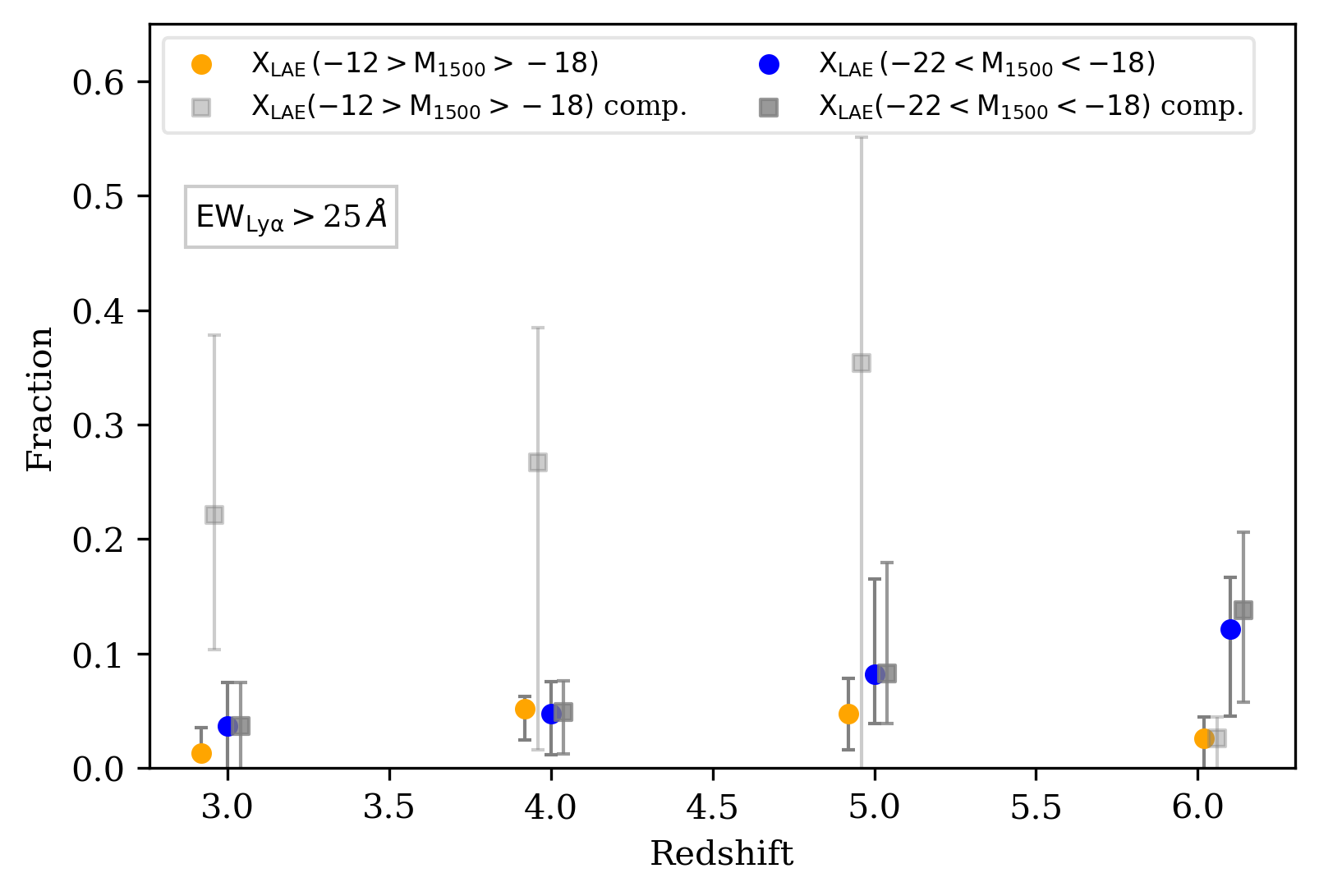}
    \\[\smallskipamount]
    \includegraphics[width=0.5\textwidth]{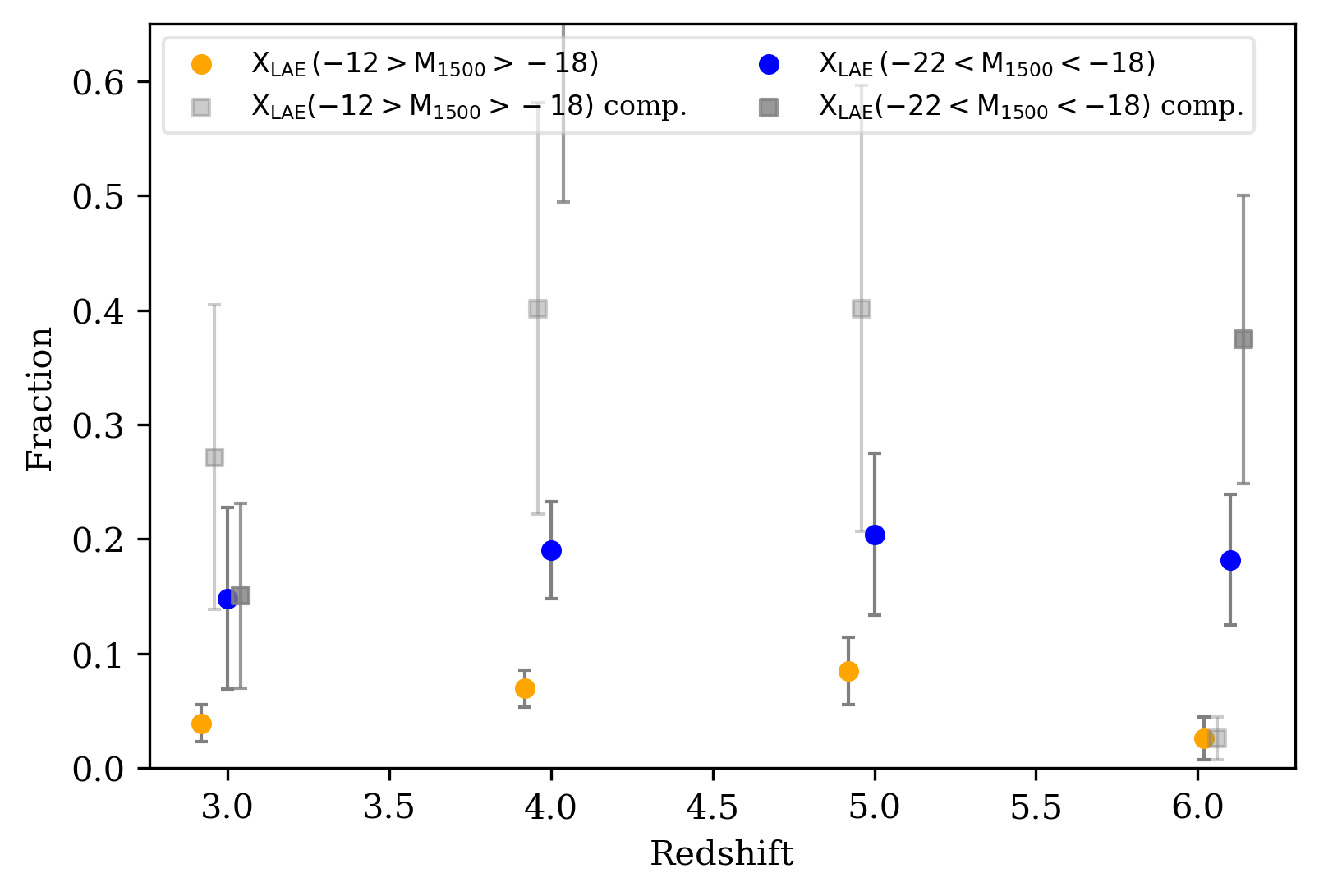}
    \caption{
    Redshift evolution of \xlae in the bright and faint populations, split at $\mathrm{M_{1500}}=-18$. The bright and faint populations are shown in blue and orange and the completeness corrections in grey scale with the bright population in darker grey. The top panel corresponds to LAEs with \ewlya $>25\,\AA$, while the lower panel shows the results when this limit is removed. The absence of the EW limit leads to a clearer separation between the two populations, suggesting that UV-bright LBGs exhibit a larger fraction of objects with \lya emission when objects with small values of \ewlya are taken into account. In the lower panel the UV-bright completeness-corrected point at $z\sim5$, omitted from the plot, is at 1.19 (see text).}
    
    \label{fig:LAEfrac18}
\end{figure}

Before discussing \xlae and its relevance to reionisation, we first outline what is typically meant by \xlae. The typical qualitative interpretation is `the fraction of UV-selected Lyman Break Galaxies that display Lyman--$\alpha$ emission'. This stems from the technique of spectroscopic follow-up on a photometric preselection of targets (e.g. \citealt{stark2010keckLAEfrac,stark2011LAEfrac}). In the case of this study, both LAE and LBG selections are performed blindly. With an IFU, we also detect LAEs that have no continuum counterpart, objects that would clearly be missed in the aforementioned type of study.\\
\indent Additionally, limits are typically placed on the inclusion of objects when calculating \xlae. The most common limits are \ewlya$>25\,\AA$ and $\mathrm{M_{1500}}>-20.25$ (\citealt{stark2011LAEfrac,schenker2014lyafracfeasibility,deBarros2017LAEfraction,mason2018LAEfrac,pentericci2018LAEfrac,arrabal2018LAEfrac}; \citetalias{GdlV2020LAEfrac} among others), however \cite{caruana2014LAEfrac} use a limit of $75\,\AA$. Other authors such as \cite{stark2010keckLAEfrac,cassata2015vimosLAEfrac,caruana2018,kusakabe2020,fuller2020LAEfrac} investigate different EW and $\mathrm{M_{1500}}$ or apparent magnitude ranges to assess the EW and UV magnitude evolution of \xlae, with the exact ranges depending on the study: the depths available and whether or not lensing is involved. In blank fields, most commonly authors investigate the bright and the faint halves of their sample, split at $\mathrm{M_{1500}}=-20.25$ (e.g. \citealt{cassata2015vimosLAEfrac,deBarros2017LAEfraction,arrabal2018LAEfrac}). Alternatively, where sample sizes allow, studies bin results in $\mathrm{M_{1500}}$ from $-22$ to $-18$ (e.g. \citealt{caruana2018,kusakabe2020}).
These are important distinctions to be aware of when comparing results from the literature. This discrepancy between the exact inclusion limits studies use, as well as the observational differences (as covered in the introduction) is likely a significant factor in the ongoing debate over the exact evolution of \xlae with redshift, \ewlya and $\mathrm{M_{1500}}$ (see Fig.~\ref{fig:LAEfrac20.25}).\\
\indent In this work we provide a perspective based on blind selections of both populations with an IFU and deep photometry with which we can assess more fully the landscape of these high-redshift galaxies. We compare the inclusion of objects only selected as LBG as well as the LAE+continuum sample described in Section~\ref{sect:matching}. Given that we detect LAEs with no continuum counterpart we also discuss the effect this has on \xlae. We investigate different $M_{1500}$ and \ewlya inclusion limits to provide context on the differing results in the literature. This last point is additionally motivated by the use of lensing fields in this study, meaning that we effectively probe a different population than blank field studies. It is therefore appropriate to investigate different regions of the \ewlya and $\mathrm{M_{1500}}$ space.\\
In Fig.~\ref{fig:LAEfrac20.25} we compare our results to results from the literature in the classical manner with inclusion limits of $\mathrm{M_{1500}}>-20.25$ and \ewlya$>25\,\AA$. The redshift bins used are $2.5<z<3.5$, $3.5<z<4.5$, $4.5<z<5.5$ and $5.5<z<7.0$. For each bin, we include four values of \xlae: the fraction of \lya emitters among LBGs (yellow dots), the fraction of \lya emitters among LBGs with the completeness corrections described in Section~\ref{sect:LAEcompleteness} (red dots) and the same two measurements including the LAE+continuum objects (yellow and red stars) (see Section~\ref{sect:matching}).\\ 
\indent Despite significant uncertainties, we see roughly the same trend as many results in the literature, with \xlae decreasing around $z=6$. The decrease of \xlae shown in our data is quicker than many equivalent measurements in the literature. If one carries forward the assumption that \xlae efficiently probes the neutral hydrogen content, this result suggests that the IGM quickly becomes neutral after $z=6$.\\
\indent The objects we detect only as LAEs (making up $\sim21\%$ of our LAE sample) would push these fractions higher, were they to be included. Considering that these objects are extremely faint in terms of UV magnitude and high in terms of \ewlya, they would make an appreciable difference in the calculation of the fraction (in a similar way to the LAE+continuum sample shown by red and yellow stars in Fig.~\ref{fig:LAEfrac20.25}). There is little reason to suspect that with deeper photometry in the region that sees the rest frame UV emission, the continuum would remain undetected for these objects.\\
\indent The high values of \xlae that we see in the first (low redshift) bin in Fig.~\ref{fig:LAEfrac20.25} are likely due to the fact that the Lyman break is harder to detect at these redshifts given the filters at our disposal. In order to check this, we perform a simulation, creating 100000 fake LBG sources using Starburst99 templates \citep{leitherer1999starburst99}. We add realistic noise to the data and spread them equally throughout the redshift and UV magnitude space probed by our study. We then attempt to select LBGs following the original selection criteria laid out in Section~\ref{sect:LBG_selection}. In the first redshift bin ($2.5<z<3.5$) the amount of LBGs selected versus the number of simulated LBGs is drastically lower than the other bins, at $<10\%$. It is worth noting that this simulation was computed for AS1063: a `best case scenario', given that AS1063 has the most filters available in the short wavelength range $(<606\,\mathrm{nm})$. For sources at redshifts between 2.5 and 3, the break would appear between $\sim320\,\mathrm{nm}$ and $\sim365\,\mathrm{nm}$. The blue filters have a shallower depth (for AS1063, F225W, F275W, F336W F390W have $5\sigma$ limiting magnitudes $\sim24.0, 25.9, 26.3, 25.4$ respectively) than the redder HST filters and due to this and their scarcity (A2744 only has F275W and F336W), we struggle to recover all the LBGs in this bin (for complete filter depth information see Table~\ref{table:filter_depths}). This can also explain why in Fig.~\ref{fig:LAELBGzm1500dist} we can see a trough in the distribution at the lower end of our redshift range, why \xlae is higher than many literature estimates in the lowest redshift bin and forms part of the reason why we see little evolution between redshifts of 3 and 5. However our results are still in agreement with the literature results at $1\sigma$. Effects like this are not unexpected considering the varying nature of LBG (and LAE) selection.\\
\indent The errors on \xlae as displayed in Figs.~\ref{fig:LAEfrac20.25} and \ref{fig:LAEfrac18} contain several components. As mentioned, there is a big impact on sample inclusion from the \ewlya limits of $25\,\AA$. Therefore it is important to quantify an error based on this. We perform 1000 Monte Carlo trials of \xlae, sorting the \ewlya values in their error bars, assumed to be Gaussian. We evolve an error estimate for \xlae based on the 16th and 84th percentile of the resulting range of results for \xlae. \\
Secondly, for the fractions to which we apply our LAE completeness correction, this correction comes with an uncertainty (see Section~\ref{sect:LAEcompleteness}) which is propagated together with the error described above and the Poissonian error count on the number of LBGs and LAEs detected. This Poissonian error in \xlae is given by: $\sigma_{P}=\left(N_{LAE}/N_{LBG}^2 + N_{LAE}^2/N_{LBG}^3\right)^{1/2}$. \\
\indent The effect of the sample inclusion limits on \ewlya are clear to see from Fig.~\ref{fig:LAEfrac20.25}, particularly in the third redshift bin. We reinforce an appreciation of this as an important matter in the calculation of \xlae and the interpretation of plots such as Fig.~\ref{fig:LAEfrac20.25}. For this study, this source of error means that it is not possible to describe a statistically significant evolution of \xlae until the highest-redshift bin. \\
\indent Similarly, the inclusion of the LAE+continuum objects in the calculation of \xlae (denoted by red and yellow stars in Fig.~\ref{fig:LAEfrac20.25}) has an appreciable effect: an average absolute difference in \xlae of 0.05 for points without completeness correction and 0.07 for those with completeness correction. This is a reflection of the impact of LBG selection in the final calculation of \xlae, particularly the SNR criterion imposed in the selection. \\
Finally, we can see from Fig.~\ref{fig:LAEfrac20.25} that the completeness correction applied to the number of LAEs plays an important role in the calculation of \xlae (see the red points in Fig.~\ref{fig:LAEfrac20.25}). Without this correction, it is also difficult to address the redshift evolution of \xlae. \\  
\indent There is interesting and unsettled debate on the relative evolution of \xlae between the UV-bright and UV-faint populations (see \citealt{stark2010keckLAEfrac,oesch2015spectroscopic,deBarros2017LAEfraction,stark2017luminouslyalpha,kusakabe2020}). In particular, this debate has consequences at higher redshift, $z\gtrsim6$, on what type of objects can be seen in \lya emission despite an increasingly neutral IGM. Relations between the UV magnitude and \lya emission at these redshifts may hold clues as to how the IGM is reionised and which objects are primarily responsible for this. In order to assess this, we split our sample in UV absolute magnitude at $\mathrm{M_{1500}}=-18$ (as this limit splits our sample roughly in half) and recalculate \xlae for the bright and faint portions. Fig.~\ref{fig:LAEfrac18} shows this comparison. We note that the $\mathrm{M_{1500}}$ ranges into which our data is split differ from many similar comparisons in the literature due to the lensed nature of our sample. We efficiently probe down to $\mathrm{M_{1500}}\sim-12$, much fainter than studies conducted in blank fields. By contrast, we have few objects as bright as $\mathrm{M_{1500}}=-22$. \\
\indent In the upper panel of Fig.~\ref{fig:LAEfrac18}, there is no statistical difference between the bright and faint halves of the sample until the final redshift bin ($5.5<z<7.0$), however we do see a rising trend in the bright half and a decreasing trend in the faint half. In order to investigate this further we remove the \ewlya limit of $25\,\AA$, resulting in the lower panel of Fig.~\ref{fig:LAEfrac18}. Based both on the result from Fig.~\ref{fig:EWvszM1500} and indications from the literature \citep{curtis2012lyafrac,stark2017luminouslyalpha}, it is likely that UV-bright galaxies exhibit \lya with small EWs: close to or below the $25\,\AA$ limit, due to spectral and spatial dispersion of the \lya photons from such systems \citep{mason2018LAEfrac}. \\
\indent This is supported by the clear distinction we see in the lower panel of Fig.~\ref{fig:LAEfrac18}. This result suggests that, while UV-fainter objects generally have higher \lya EWs, the LAE fraction is greater among UV-brighter objects, albeit with those objects exhibiting smaller \ewlya. Our findings are especially interesting in the context of the number of low \ewlya, UV-bright objects reported towards and into the epoch of reionisation (\citealt{curtis2012lyafrac,oesch2015spectroscopic,stark2017luminouslyalpha,mason2018brightlya,mason2018LAEfrac,larson2022reionisation,bunker2023GNz11,Witten2023lya_inreionisation}, see also \citealt{ando2006stronglyadef}). This result is often taken as surprising given that the increasingly neutral IGM should block much of the \lya emission at redshifts greater than 6. However, it is possible that with an increasingly neutral IGM, the more luminous galaxies ionise a surrounding bubble, making it easier for \lya emission to escape. These galaxies are also more likely to be situated in reionised overdensities \citep{Matthee2015brightlyareionise,mason2018brightlya,Witten2023lya_inreionisation}. \citet{Witten2023lya_inreionisation}, cataloguing 8 LAEs within the epoch of reionisation, most of which have small \ewlya ($<25\,\AA$), also suggest that mergers and tidal interactions with neighbours may be responsible for enhanced \lya visibility during the epoch of reionisation. Reionised bubbles created by these processes and these bright galaxies would allow the \lya photons to redshift out of resonance by the time they encounter significant neutral hydrogen, and hence would not be absorbed or scattered.\\
\indent More detailed, individual analysis of individual objects would be necessary to firmly support these conclusions for our high-redshift objects but the difference seen in the highest redshift bins of the graphs in Fig.~\ref{fig:LAEfrac18} broadly supports the enhanced visibility of UV-bright, low-\ewlya LAEs in the epoch of reionisation.\\
\indent We caution the reader that we detect a population of high-EW LAEs undetected in the continuum, and if these objects were to be detected with deeper photometry, they could modify this result as they would be UV-faint with high \ewlya. However, they are roughly equally spread over the redshift range probed so may leave the trends in Fig.~\ref{fig:LAEfrac18} unchanged. \\
\indent We plot the completeness corrected results in Fig.~\ref{fig:LAEfrac18} in grey scale, darker grey representing the UV-bright portion of the sample and lighter grey representing the UV-faint portion. Due to the smaller sample sizes in these magnitude-split bins, these corrections have large errors where one or a few LAEs with a small completeness values, accounting for many completeness-corrected LAEs, can dominate the determination of \xlae for a given bin. The lower panel of Fig.~\ref{fig:LAEfrac18} provides an example: \xlae for the bright half of the sample in the bin $4.5<z<5.5$ is at 1.19 i.e. accounting for the completeness correction, there are more UV-bright LAEs than LBGs in this bin. Noting that this is caused by individual highly incomplete objects, we do not interpret this as meaningful in the context of \xlae with respect to the evolution of the IGM and reionisation. On the other hand, we note that the completeness-corrected trends also broadly support the conclusions outlined earlier in this section, particularly in the highest-redshift bins, where the difference between the UV-bright and UV-faint portions of the sample is even more pronounced.

\section{Conclusions}
\label{sect:conclusion}
 We present an assessment of the interrelation of the faint \lya Emitter and Lyman Break Galaxy population viewed with MUSE IFU spectroscopy and deep HST photometry. We can access faint populations unseen in blank fields through the magnification provided by four lensing clusters, pushing LBG detections down to $\mathrm{M_{1500}}\sim-12$ and LAE detections down to \lya luminosities of $\mathrm{log(L_{Ly\alpha})\sim39\,erg\,s^{-1}}$. We find LAEs with no detected continuum counterpart, actors that play an increasingly important role in the regime fainter than $M_{1500}\sim-16$. We summarise our main results as such:\\
\indent -- Our results for \xlae agree with findings in the literature and if we accept that it is unlikely that the evolution of one or both populations considered makes a significant difference to \xlae, our faint sample supports the conclusions of brighter studies, that the Universe is reionising at $z=6$ and beyond. However based on the results of this study, we find little to no evolution between redshifts of 3 and 5. In part this is likely due to the greater effect of LBG selection incompleteness in the lower redshift regions. The scatter in this redshift range in the literature is likely due to issues related to the LBG selection and population completeness as discussed in Section~\ref{sect:LAEfraction}.\\
\indent -- We compare the evolution of \xlae for the bright and faint halves of our sample, split at $\mathrm{M_{1500}}=-18$. We find a different trend between the two; \xlae for the bright half rises towards $z=6$ and \xlae for the faint half falls. The difference in \xlae between the UV-bright and UV-faint half is statistically significant in the highest-redshift bin, indicating that this effect increases as the Universe becomes more neutral.\\
\indent -- To further investigate this, we remove the $25\,\AA$ limit in the calculation of \xlae. Without this limit, a clear distinction is seen between the bright and faint halves of the sample, suggesting that while high-EW LAEs tend to be UV-faint, there is a population of low-EW LAEs among the UV-bright population. This can account for some of the newly detected, low-EW, UV-bright objects around and in the epoch of reionisation. \\
\indent -- When considering the UV properties of our faint samples, we see the typical picture of LAEs as high-SFR, low-dust, UV-faint galaxies, extending previously observed trends to very faint UV magnitudes and \lya luminosities. The strongest trend is seen when comparing \ewlya to $\mathrm{M_{1500}}$ (Fig.~\ref{fig:EWvszM1500}). UV-fainter objects on average display larger \ewlya, a trend which likely continues down to $\mathrm{M_{1500}}\sim-12$ based on our sample.\\
\indent -- We also extend, to fainter luminosities and with a larger sample, the trend that UV-brighter galaxies tend to exhibit greater \fuv than \flya, whereas UV-fainter galaxies are distributed more towards the \fuv = \flya line, sometimes even displaying \flya > \fuv. This has been attributed to UV-bright populations being older, dustier and more chemically evolved, decreasing their \lya emission (\citealt{ando2006stronglyadef}; \citetalias{GdlV2020LAEfrac}). This is tentatively reported for our sample, particularly as there are significant uncertainties associated with the \sfruv values of the fainter galaxies.\\
\indent -- Even with the deepest HST photometry available, the presence of continuum-undetected LAEs remains important at UV magnitudes fainter than $\mathrm{M_{1500}}=-16$. Surveys with JWST can help to shed light on the effect of much deeper photometry, albeit at different wavelengths than the HST.


\begin{acknowledgements}
This work is done based on observations made with ESO Telescopes
at the La Silla Paranal Observatory under programme IDs 060.A-9345, 092.A-0472,
094.A-0115, 095.A-0181, 096.A-0710, 097.A0269, 100.A-0249, and
294.A-5032. Also based on observations obtained with the NASA/ESA Hubble
Space Telescope, retrieved from the Mikulski Archive for Space Telescopes
(MAST) at the Space Telescope Science Institute (STScI). STScI is operated by
the Association of Universities for Research in Astronomy, Inc. under NASA
contract NAS 5-26555. 
All plots in this paper were created using Matplotlib (Hunter 2007). 
Part of this work was supported by the French CNRS, the Aix-Marseille University, the French Programme National de Cosmologie et Galaxies (PNCG) of CNRS/INSU with INP and IN2P3, co-funded by CEA and CNES.
This work also received support from the French government under the France 2030 investment plan, as part of the Excellence Initiative of Aix-Marseille University - A*MIDEX (AMX-19-IET-008 - IPhU).
Financial support from the World Laboratory, the Odon Vallet Foundation and VNSC is gratefully acknowledged. Tran Thi Thai was funded by Vingroup JSC and supported by the Master, PhD Scholarship Programme of Vingroup Innovation Foundation (VINIF), Institute of Big Data, code VINIF.2022.TS.107.        
\end{acknowledgements}

\bibliographystyle{aa} 
\bibliography{bib.bib} 

\begin{thebibliography}{103}
\expandafter\ifx\csname natexlab\endcsname\relax\def\natexlab#1{#1}\fi

\bibitem[{{Acebron} {et~al.}(2018){Acebron}, {Cibirka}, {Zitrin}, {Coe},
  {Agulli}, {Sharon}, {Brada{\v{c}}}, {Frye}, {Livermore}, {Mahler}, {Salmon},
  {Umetsu}, {Bradley}, {Andrade-Santos}, {Avila}, {Carrasco}, {Cerny},
  {Czakon}, {Dawson}, {Hoag}, {Huang}, {Johnson}, {Jones}, {Kikuchihara},
  {Lam}, {Lovisari}, {Mainali}, {Oesch}, {Ogaz}, {Ouchi}, {Past},
  {Paterno-Mahler}, {Peterson}, {Ryan}, {Sendra-Server}, {Stark}, {Strait},
  {Toft}, {Trenti}, \& {Vulcani}}]{Acebron2018strong_lensing_analysis}
{Acebron}, A., {Cibirka}, N., {Zitrin}, A., {et~al.} 2018, \apj, 858, 42

\bibitem[{{Acebron} {et~al.}(2017){Acebron}, {Jullo}, {Limousin}, {Tilquin},
  {Giocoli}, {Jauzac}, {Mahler}, \& {Richard}}]{Acebron2017lensing_systematics}
{Acebron}, A., {Jullo}, E., {Limousin}, M., {et~al.} 2017, \mnras, 470, 1809

\bibitem[{Ando {et~al.}(2006)Ando, Ohta, Iwata, Akiyama, Aoki, \&
  Tamura}]{ando2006stronglyadef}
Ando, M., Ohta, K., Iwata, I., {et~al.} 2006, The Astrophysical Journal, 645,
  L9

\bibitem[{Arrabal~Haro {et~al.}(2018)Arrabal~Haro, Rodr{\'\i}guez~Espinosa,
  Mu{\~n}oz-Tu{\~n}{\'o}n, P{\'e}rez-Gonz{\'a}lez, Dannerbauer, Bongiovanni,
  Barro, Cava, Lumbreras-Calle, Hern{\'a}n-Caballero,
  {et~al.}}]{arrabal2018LAEfrac}
Arrabal~Haro, P., Rodr{\'\i}guez~Espinosa, J., Mu{\~n}oz-Tu{\~n}{\'o}n, C.,
  {et~al.} 2018, Monthly Notices of the Royal Astronomical Society, 478, 3740

\bibitem[{Bacon {et~al.}(2010)Bacon, Accardo, Adjali, Anwand, Bauer, Biswas,
  Blaizot, Boudon, Brau-Nogue, Brinchmann, Caillier, Capoani, Carollo, Contini,
  Couderc, Daguis{\'e}, Deiries, Delabre, Dreizler, Dubois, Dupieux, Dupuy,
  Emsellem, Fechner, Fleischmann, Fran{\c{c}}ois, Gallou, Gharsa, Glindemann,
  Gojak, Guiderdoni, Hansali, Hahn, Jarno, Kelz, Koehler, Kosmalski, Laurent,
  Floch, Lilly, Lizon, Loupias, Manescau, Monstein, Nicklas, Olaya, Pares,
  Pasquini, P{\'e}contal-Rousset, Pell{\'o}, Petit, Popow, Reiss, Remillieux,
  Renault, Roth, Rupprecht, Serre, Schaye, Soucail, Steinmetz, Streicher,
  Stuik, H, Vernet, Weilbacher, Wisotzki, \& Yerle}]{Bacon_et_al_2010_SPIE}
Bacon, R., Accardo, M., Adjali, L., {et~al.} 2010, in Ground-based and Airborne
  Instrumentation for Astronomy III, ed. I.~S. McLean, S.~K. Ramsay, \&
  H.~Takami, Vol. 7735, International Society for Optics and Photonics (SPIE),
  773508

\bibitem[{Bacon {et~al.}(2023)Bacon, Brinchmann, Conseil, Maseda, Nanayakkara,
  Wendt, Bacher, Mary, Weilbacher, Krajnovi{\'c},
  {et~al.}}]{bacon2022musedatareleaseII}
Bacon, R., Brinchmann, J., Conseil, S., {et~al.} 2023, Astronomy and
  Astrophysics, 670, A4

\bibitem[{Beauchesne {et~al.}(2023)Beauchesne, Cl{\'e}ment, Hibon, Limousin,
  Eckert, Kneib, Richard, Natarajan, Jauzac, Montes,
  {et~al.}}]{beauchesne2023AS1063lensmodel}
Beauchesne, B., Cl{\'e}ment, B., Hibon, P., {et~al.} 2023, arXiv preprint
  arXiv:2301.10907

\bibitem[{Bertin \& Arnouts(1996)}]{EB96SEx}
Bertin, E. \& Arnouts, S. 1996, Astronomy and astrophysics supplement series,
  117, 393

\bibitem[{Bina {et~al.}(2016)Bina, Pell{\'o}, Richard, Lewis, Patr{\'\i}cio,
  Cantalupo, Herenz, Soto, Weilbacher, Bacon, {et~al.}}]{bina2016muse}
Bina, D., Pell{\'o}, R., Richard, J., {et~al.} 2016, Astronomy \& Astrophysics,
  590, A14

\bibitem[{Bolan {et~al.}(2022)Bolan, Lemaux, Mason, Brada{\v{c}}, Treu, Strait,
  Pelliccia, Pentericci, \& Malkan}]{bolan2022neutralfraction}
Bolan, P., Lemaux, B.~C., Mason, C., {et~al.} 2022, Monthly Notices of the
  Royal Astronomical Society, 517, 3263

\bibitem[{Bolton \& Haehnelt(2013)}]{bolton2013IGMlya_alternate}
Bolton, J.~S. \& Haehnelt, M.~G. 2013, Monthly Notices of the Royal
  Astronomical Society, 429, 1695

\bibitem[{Bolzonella {et~al.}(2000)Bolzonella, Miralles, \&
  Pell{\'o}}]{MB00hyperz}
Bolzonella, M., Miralles, J.-M., \& Pell{\'o}, R. 2000, Astronomy and
  Astrophysics, 363, 476

\bibitem[{Bouwens {et~al.}(2021)Bouwens, Oesch, Stefanon, Illingworth,
  Labb{\'e}, Reddy, Atek, Montes, Naidu, Nanayakkara,
  {et~al.}}]{bouwens2021UVLF}
Bouwens, R., Oesch, P., Stefanon, M., {et~al.} 2021, The Astronomical Journal,
  162, 47

\bibitem[{{Bouwens} {et~al.}(2022){Bouwens}, {Illingworth}, {Ellis}, {Oesch},
  \& {Stefanon}}]{Bouwens2022UVLF2z9}
{Bouwens}, R.~J., {Illingworth}, G., {Ellis}, R.~S., {Oesch}, P., \&
  {Stefanon}, M. 2022, \apj, 940, 55

\bibitem[{Bouwens {et~al.}(2015{\natexlab{a}})Bouwens, Illingworth, Oesch,
  Trenti, Labb{\'e}, Bradley, Carollo, Van~Dokkum, Gonzalez, Holwerda,
  {et~al.}}]{bouwens2015UVLF}
Bouwens, R.~J., Illingworth, G., Oesch, P., {et~al.} 2015{\natexlab{a}}, The
  Astrophysical Journal, 803, 34

\bibitem[{Bouwens {et~al.}(2015{\natexlab{b}})Bouwens, Illingworth, Oesch,
  Caruana, Holwerda, Smit, \& Wilkins}]{Bouwens2015planckreionisation}
Bouwens, R.~J., Illingworth, G.~D., Oesch, P.~A., {et~al.} 2015{\natexlab{b}},
  The Astrophysical Journal, 811, 140

\bibitem[{Brammer {et~al.}(2016)Brammer, Marchesini, Labbé, Spitler,
  Lange-Vagle, Barker, Tanaka, Fontana, Galametz, Ferré-Mateu, Kodama,
  Lundgren, Martis, Muzzin, Stefanon, Toft, van~der Wel, Vulcani, \&
  Whitaker}]{Brammer2016KeckHFF}
Brammer, G.~B., Marchesini, D., Labbé, I., {et~al.} 2016, The Astrophysical
  Journal Supplement Series, 226, 6

\bibitem[{Bruzual \& Charlot(2003)}]{BandC03SED}
Bruzual, G. \& Charlot, S. 2003, Monthly Notices of the Royal Astronomical
  Society, 344, 1000

\bibitem[{Bunker {et~al.}(2023)Bunker, Saxena, Cameron, Willott, Curtis-Lake,
  Jakobsen, Carniani, Smit, Maiolino, Witstok, {et~al.}}]{bunker2023GNz11}
Bunker, A.~J., Saxena, A., Cameron, A.~J., {et~al.} 2023, arXiv preprint
  arXiv:2302.07256

\bibitem[{Calzetti {et~al.}(2000)Calzetti, Armus, Bohlin, Kinney, Koornneef, \&
  Storchi-Bergmann}]{Calzetti00dust}
Calzetti, D., Armus, L., Bohlin, R.~C., {et~al.} 2000, The Astrophysical
  Journal, 533, 682

\bibitem[{Caruana {et~al.}(2014)Caruana, Bunker, Wilkins, Stanway, Lorenzoni,
  Jarvis, \& Ebert}]{caruana2014LAEfrac}
Caruana, J., Bunker, A.~J., Wilkins, S.~M., {et~al.} 2014, Monthly Notices of
  the Royal Astronomical Society, 443, 2831

\bibitem[{Caruana {et~al.}(2018)Caruana, Wisotzki, Herenz, Kerutt, Urrutia,
  Schmidt, Bouwens, Brinchmann, Cantalupo, Carollo, {et~al.}}]{caruana2018}
Caruana, J., Wisotzki, L., Herenz, E.~C., {et~al.} 2018, Monthly Notices of the
  Royal Astronomical Society, 473, 30

\bibitem[{Cassata {et~al.}(2015)Cassata, Tasca, Le~F{\`e}vre, Lemaux, Garilli,
  Le~Brun, Maccagni, Pentericci, Thomas, Vanzella,
  {et~al.}}]{cassata2015vimosLAEfrac}
Cassata, P., Tasca, L., Le~F{\`e}vre, O., {et~al.} 2015, Astronomy \&
  Astrophysics, 573, A24

\bibitem[{Castellano {et~al.}(2012)Castellano, Fontana, Grazian, Pentericci,
  Santini, Koekemoer, Cristiani, Galametz, Gallerani, Vanzella,
  {et~al.}}]{castellano2012Beta}
Castellano, M., Fontana, A., Grazian, A., {et~al.} 2012, Astronomy \&
  Astrophysics, 540, A39

\bibitem[{Claeyssens {et~al.}(2022)Claeyssens, Richard, Blaizot, Garel,
  Kusakabe, Bacon, Bauer, Guaita, Jeanneau, Lagattuta, {et~al.}}]{AC2022LLAMAS}
Claeyssens, A., Richard, J., Blaizot, J., {et~al.} 2022, Astronomy and
  Astrophysics, 666, A78

\bibitem[{Coleman {et~al.}(1980)Coleman, Wu, \& Weedman}]{CWW1980colors}
Coleman, G., Wu, C.-C., \& Weedman, D. 1980, Astrophysical Journal Supplement
  Series, vol. 43, July 1980, p. 393-416. Research sponsored by the Koninklijke
  Nederlandse Akademie van Wetenschappen;, 43, 393

\bibitem[{Curtis-Lake {et~al.}(2012)Curtis-Lake, McLure, Pearce, Dunlop,
  Cirasuolo, Stark, Almaini, Bradshaw, Chuter, Foucaud,
  {et~al.}}]{curtis2012lyafrac}
Curtis-Lake, E., McLure, R., Pearce, H., {et~al.} 2012, Monthly Notices of the
  Royal Astronomical Society, 422, 1425

\bibitem[{Dayal {et~al.}(2011)Dayal, Maselli, \&
  Ferrara}]{dayal2011lyavisibility}
Dayal, P., Maselli, A., \& Ferrara, A. 2011, Monthly Notices of the Royal
  Astronomical Society, 410, 830

\bibitem[{De~Barros {et~al.}(2017)De~Barros, Pentericci, Vanzella, Castellano,
  Fontana, Grazian, Conselice, Yan, Koekemoer, Cristiani,
  {et~al.}}]{deBarros2017LAEfraction}
De~Barros, S., Pentericci, L., Vanzella, E., {et~al.} 2017, Astronomy \&
  Astrophysics, 608, A123

\bibitem[{de~La~Vieuville {et~al.}(2019)de~La~Vieuville, Bina, Pello, Mahler,
  Richard, Drake, Herenz, Bauer, Cl{\'e}ment, Lagattuta,
  {et~al.}}]{GdlV2019LAELF}
de~La~Vieuville, G., Bina, D., Pello, R., {et~al.} 2019, Astronomy \&
  Astrophysics, 628, A3

\bibitem[{de~La~Vieuville {et~al.}(2020)de~La~Vieuville, Pell{\'o}, Richard,
  Mahler, L{\'e}v{\^e}que, Bauer, Lagattuta, Blaizot, Contini, Guaita,
  {et~al.}}]{GdlV2020LAEfrac}
de~La~Vieuville, G., Pell{\'o}, R., Richard, J., {et~al.} 2020, Astronomy \&
  Astrophysics, 644, A39

\bibitem[{Dijkstra(2016)}]{dijkstra2016LAEreview}
Dijkstra, M. 2016, Understanding the Epoch of Cosmic Reionization: Challenges
  and Progress, 145

\bibitem[{Dijkstra {et~al.}(2011)Dijkstra, Mesinger, \&
  Wyithe}]{dijkstra2011lyadetectability}
Dijkstra, M., Mesinger, A., \& Wyithe, J. S.~B. 2011, Monthly Notices of the
  Royal Astronomical Society, 414, 2139

\bibitem[{Drake {et~al.}(2017)Drake, Garel, Wisotzki, Leclercq, Hashimoto,
  Richard, Bacon, Blaizot, Caruana, Conseil, {et~al.}}]{drake2017MUSELAELF}
Drake, A.-B., Garel, T., Wisotzki, L., {et~al.} 2017, Astronomy \&
  Astrophysics, 608, A6

\bibitem[{Fan {et~al.}(2006)Fan, Strauss, Becker, White, Gunn, Knapp, Richards,
  Schneider, Brinkmann, \& Fukugita}]{fan2006reionisation}
Fan, X., Strauss, M.~A., Becker, R.~H., {et~al.} 2006, The Astronomical
  Journal, 132, 117

\bibitem[{Finkelstein {et~al.}(2015)Finkelstein, Ryan, Papovich, Dickinson,
  Song, Somerville, Ferguson, Salmon, Giavalisco, Koekemoer, Ashby, Behroozi,
  Castellano, Dunlop, Faber, Fazio, Fontana, Grogin, Hathi, Jaacks, Kocevski,
  Livermore, McLure, Merlin, Mobasher, Newman, Rafelski, Tilvi, \&
  Willner}]{Finkelstein_2015}
Finkelstein, S.~L., Ryan, R.~E., Papovich, C., {et~al.} 2015, The Astrophysical
  Journal, 810, 71

\bibitem[{Foreman-Mackey {et~al.}(2013)Foreman-Mackey, Hogg, Lang, \&
  Goodman}]{foreman2013emcee}
Foreman-Mackey, D., Hogg, D.~W., Lang, D., \& Goodman, J. 2013, Publications of
  the Astronomical Society of the Pacific, 125, 306

\bibitem[{Fuller {et~al.}(2020)Fuller, Lemaux, Brada{\v{c}}, Hoag, Schmidt,
  Huang, Strait, Mason, Treu, Pentericci, {et~al.}}]{fuller2020LAEfrac}
Fuller, S., Lemaux, B., Brada{\v{c}}, M., {et~al.} 2020, The Astrophysical
  Journal, 896, 156

\bibitem[{{Furtak} {et~al.}(2021){Furtak}, {Atek}, {Lehnert}, {Chevallard}, \&
  {Charlot}}]{Furtak2021cluster_stellasmassfunc}
{Furtak}, L.~J., {Atek}, H., {Lehnert}, M.~D., {Chevallard}, J., \& {Charlot},
  S. 2021, \mnras, 501, 1568

\bibitem[{Grazian {et~al.}(2018)Grazian, Giallongo, Boutsia, Cristiani,
  Vanzella, Scarlata, Santini, Pentericci, Merlin, Menci,
  {et~al.}}]{Grazian2018AGN}
Grazian, A., Giallongo, E., Boutsia, K., {et~al.} 2018, Astronomy \&
  Astrophysics, 613, A44

\bibitem[{Gronke {et~al.}(2021)Gronke, Ocvirk, Mason, Matthee, Bosman, Sorce,
  Lewis, Ahn, Aubert, Dawoodbhoy, {et~al.}}]{gronke2021lyatransmission}
Gronke, M., Ocvirk, P., Mason, C., {et~al.} 2021, Monthly Notices of the Royal
  Astronomical Society, 508, 3697

\bibitem[{Hayes {et~al.}(2013)Hayes, {\"O}stlin, Schaerer, Verhamme, Mas-Hesse,
  Adamo, Atek, Cannon, Duval, Guaita, {et~al.}}]{hayes2013lyadust}
Hayes, M., {\"O}stlin, G., Schaerer, D., {et~al.} 2013, The Astrophysical
  Journal Letters, 765, L27

\bibitem[{Hayes {et~al.}(2011)Hayes, Schaerer, {\"O}stlin, Mas-Hesse, Atek, \&
  Kunth}]{hayes2011redshiftevofdust}
Hayes, M., Schaerer, D., {\"O}stlin, G., {et~al.} 2011, The Astrophysical
  Journal, 730, 8

\bibitem[{Herenz {et~al.}(2019)Herenz, Wisotzki, Saust, Kerutt, Urrutia,
  Diener, Schmidt, Marino, De~la Vieuville, Boogaard,
  {et~al.}}]{herenz2019MUSELAELF}
Herenz, E.~C., Wisotzki, L., Saust, R., {et~al.} 2019, Astronomy \&
  Astrophysics, 621, A107

\bibitem[{Hoag {et~al.}(2019{\natexlab{a}})Hoag, Brada{\v{c}}, Huang, Mason,
  Treu, Schmidt, Trenti, Strait, Lemaux, Finney,
  {et~al.}}]{hoag2019reionisation_z=7.7}
Hoag, A., Brada{\v{c}}, M., Huang, K., {et~al.} 2019{\natexlab{a}}, The
  Astrophysical Journal, 878, 12

\bibitem[{Hoag {et~al.}(2019{\natexlab{b}})Hoag, Treu, Pentericci, Amorin,
  Bolzonella, Brada{\v{c}}, Castellano, Cullen, Fynbo, Garilli,
  {et~al.}}]{hoag2019lyaoffset}
Hoag, A., Treu, T., Pentericci, L., {et~al.} 2019{\natexlab{b}}, Monthly
  Notices of the Royal Astronomical Society, 488, 706

\bibitem[{Hutter {et~al.}(2014)Hutter, Dayal, Partl, \&
  M{\"u}ller}]{hutter2014lyavisibility}
Hutter, A., Dayal, P., Partl, A.~M., \& M{\"u}ller, V. 2014, Monthly Notices of
  the Royal Astronomical Society, 441, 2861

\bibitem[{{Jiang} {et~al.}(2022){Jiang}, {Ning}, {Fan}, {Ho}, {Luo}, {Wang},
  {Wu}, {Wu}, {Yang}, \& {Zheng}}]{JiangAGN2022}
{Jiang}, L., {Ning}, Y., {Fan}, X., {et~al.} 2022, Nature Astronomy, 6, 850

\bibitem[{Jullo \& Kneib(2009)}]{jullokneib2009lenstool2}
Jullo, E. \& Kneib, J.-P. 2009, Monthly Notices of the Royal Astronomical
  Society, 395, 1319

\bibitem[{Jullo {et~al.}(2007)Jullo, Kneib, Limousin, Eliasdottir, Marshall, \&
  Verdugo}]{jullo2007lenstool}
Jullo, E., Kneib, J.-P., Limousin, M., {et~al.} 2007, New Journal of Physics,
  9, 447

\bibitem[{Kakiichi {et~al.}(2016)Kakiichi, Dijkstra, Ciardi, \&
  Graziani}]{kakiichi2016lyamorphology_reionisation}
Kakiichi, K., Dijkstra, M., Ciardi, B., \& Graziani, L. 2016, Monthly Notices
  of the Royal Astronomical Society, 463, 4019

\bibitem[{Kennicutt~Jr(1998)}]{kennicutt1998Schmidtlaw}
Kennicutt~Jr, R.~C. 1998, The astrophysical journal, 498, 541

\bibitem[{Kneib {et~al.}(2011)Kneib, Bonnet, Golse, Sand, Jullo, \&
  Marshall}]{kneibnew2011lenstool}
Kneib, J.-P., Bonnet, H., Golse, G., {et~al.} 2011, Astrophysics Source Code
  Library, ascl

\bibitem[{Kneib {et~al.}(1996)Kneib, Ellis, Smail, Couch, \&
  Sharples}]{kneib1996lenstooloriginal}
Kneib, J.-P., Ellis, R.~S., Smail, I., Couch, W., \& Sharples, R. 1996, The
  Astrophysical Journal, 471, 643

\bibitem[{Kusakabe {et~al.}(2020)Kusakabe, Blaizot, Garel, Verhamme, Bacon,
  Richard, Hashimoto, Inami, Conseil, Guiderdoni, {et~al.}}]{kusakabe2020}
Kusakabe, H., Blaizot, J., Garel, T., {et~al.} 2020, Astronomy \& Astrophysics,
  638, A12

\bibitem[{Larson {et~al.}(2022)Larson, Finkelstein, Hutchison, Papovich,
  Bagley, Dickinson, Rojas-Ruiz, Ferguson, Jung, Giavalisco,
  {et~al.}}]{larson2022reionisation}
Larson, R.~L., Finkelstein, S.~L., Hutchison, T.~A., {et~al.} 2022, The
  Astrophysical Journal, 930, 104

\bibitem[{Leclercq {et~al.}(2017)Leclercq, Bacon, Wisotzki, Mitchell, Garel,
  Verhamme, Blaizot, Hashimoto, Herenz, Conseil,
  {et~al.}}]{leclercq2017lyahaloes}
Leclercq, F., Bacon, R., Wisotzki, L., {et~al.} 2017, Astronomy \&
  Astrophysics, 608, A8

\bibitem[{Leitherer {et~al.}(1999)Leitherer, Schaerer, Goldader, Delgado,
  Robert, Kune, de~Mello, Devost, \& Heckman}]{leitherer1999starburst99}
Leitherer, C., Schaerer, D., Goldader, J.~D., {et~al.} 1999, The Astrophysical
  Journal Supplement Series, 123, 3

\bibitem[{Leonova {et~al.}(2022)Leonova, Oesch, Qin, Naidu, Wyithe, De~Barros,
  Bouwens, Ellis, Endsley, Hutter, {et~al.}}]{leonova2022prevalence}
Leonova, E., Oesch, P., Qin, Y., {et~al.} 2022, Monthly Notices of the Royal
  Astronomical Society, 515, 5790

\bibitem[{Livermore {et~al.}(2017)Livermore, Finkelstein, \&
  Lotz}]{Livermore_2017}
Livermore, R.~C., Finkelstein, S.~L., \& Lotz, J.~M. 2017, The Astrophysical
  Journal, 835, 113

\bibitem[{Lotz {et~al.}(2017)Lotz, Koekemoer, Coe, Grogin, Capak, Mack,
  Anderson, Avila, Barker, Borncamp, {et~al.}}]{lotz2017frontier}
Lotz, J.~e., Koekemoer, A., Coe, D., {et~al.} 2017, The Astrophysical Journal,
  837, 97

\bibitem[{Lu {et~al.}(2022)Lu, Goto, Hashimoto, Santos, Wong, Kim, Hsiao,
  Kilerci, Ho, Nagao, {et~al.}}]{lu2022subarureionisation}
Lu, T.-Y., Goto, T., Hashimoto, T., {et~al.} 2022, Monthly Notices of the Royal
  Astronomical Society, 517, 1264

\bibitem[{Madau \& Haardt(2015)}]{MadauHardtAGN_2015}
Madau, P. \& Haardt, F. 2015, The Astrophysical Journal Letters, 813, L8

\bibitem[{Mason {et~al.}(2018{\natexlab{a}})Mason, Treu, De~Barros, Dijkstra,
  Fontana, Mesinger, Pentericci, Trenti, \& Vanzella}]{mason2018brightlya}
Mason, C.~A., Treu, T., De~Barros, S., {et~al.} 2018{\natexlab{a}}, The
  Astrophysical Journal Letters, 857, L11

\bibitem[{Mason {et~al.}(2018{\natexlab{b}})Mason, Treu, Dijkstra, Mesinger,
  Trenti, Pentericci, De~Barros, \& Vanzella}]{mason2018LAEfrac}
Mason, C.~A., Treu, T., Dijkstra, M., {et~al.} 2018{\natexlab{b}}, The
  Astrophysical Journal, 856, 2

\bibitem[{Matthee {et~al.}(2022)Matthee, Naidu, Pezzulli, Gronke, Sobral,
  Oesch, Hayes, Erb, Schaerer, Amor{\'\i}n, {et~al.}}]{Mathee2022brightlya}
Matthee, J., Naidu, R.~P., Pezzulli, G., {et~al.} 2022, Monthly Notices of the
  Royal Astronomical Society, 512, 5960

\bibitem[{Matthee {et~al.}(2016)Matthee, Sobral, Oteo, Best, Smail,
  R{\"o}ttgering, \& Paulino-Afonso}]{matthee2016calymha_lyaesc}
Matthee, J., Sobral, D., Oteo, I., {et~al.} 2016, Monthly Notices of the Royal
  Astronomical Society, 458, 449

\bibitem[{{Matthee} {et~al.}(2015){Matthee}, {Sobral}, {Santos},
  {R{\"o}ttgering}, {Darvish}, \& {Mobasher}}]{Matthee2015brightlyareionise}
{Matthee}, J., {Sobral}, D., {Santos}, S., {et~al.} 2015, \mnras, 451, 400

\bibitem[{McGreer {et~al.}(2018)McGreer, Fan, Jiang, \& Cai}]{Mcgreer2018AGN}
McGreer, I.~D., Fan, X., Jiang, L., \& Cai, Z. 2018, The Astronomical Journal,
  155, 131

\bibitem[{McGreer {et~al.}(2015)McGreer, Mesinger, \&
  D'Odorico}]{mcgreer2015reionisation}
McGreer, I.~D., Mesinger, A., \& D'Odorico, V. 2015, Monthly Notices of the
  Royal Astronomical Society, 447, 499

\bibitem[{{Meneghetti} {et~al.}(2017){Meneghetti}, {Natarajan}, {Coe},
  {Contini}, {De Lucia}, {Giocoli}, {Acebron}, {Borgani}, {Bradac}, {Diego},
  {Hoag}, {Ishigaki}, {Johnson}, {Jullo}, {Kawamata}, {Lam}, {Limousin},
  {Liesenborgs}, {Oguri}, {Sebesta}, {Sharon}, {Williams}, \&
  {Zitrin}}]{Meneghetti2017HFF}
{Meneghetti}, M., {Natarajan}, P., {Coe}, D., {et~al.} 2017, \mnras, 472, 3177

\bibitem[{Mesinger {et~al.}(2015)Mesinger, Aykutalp, Vanzella, Pentericci,
  Ferrara, \& Dijkstra}]{mesinger2015IGMlya}
Mesinger, A., Aykutalp, A., Vanzella, E., {et~al.} 2015, Monthly Notices of the
  Royal Astronomical Society, 446, 566

\bibitem[{Naidu {et~al.}(2020)Naidu, Tacchella, Mason, Bose, Oesch, \&
  Conroy}]{naidu2020brightlya}
Naidu, R.~P., Tacchella, S., Mason, C.~A., {et~al.} 2020, The Astrophysical
  Journal, 892, 109

\bibitem[{Oesch {et~al.}(2015)Oesch, Van~Dokkum, Illingworth, Bouwens,
  Momcheva, Holden, Roberts-Borsani, Smit, Franx, Labb{\'e},
  {et~al.}}]{oesch2015spectroscopic}
Oesch, P., Van~Dokkum, P., Illingworth, G., {et~al.} 2015, The Astrophysical
  Journal Letters, 804, L30

\bibitem[{{Oke} \& {Gunn}(1983)}]{OkeGunn1983}
{Oke}, J.~B. \& {Gunn}, J.~E. 1983, \apj, 266, 713

\bibitem[{Onoue {et~al.}(2017)Onoue, Kashikawa, Willott, Hibon, Im, Furusawa,
  Harikane, Imanishi, Ishikawa, Kikuta, {et~al.}}]{Onoue2017AGN}
Onoue, M., Kashikawa, N., Willott, C.~J., {et~al.} 2017, The Astrophysical
  Journal Letters, 847, L15

\bibitem[{Parsa {et~al.}(2018)Parsa, Dunlop, \&
  McLure}]{parsa2018noAGNcontribution}
Parsa, S., Dunlop, J.~S., \& McLure, R.~J. 2018, Monthly Notices of the Royal
  Astronomical Society, 474, 2904

\bibitem[{Pentericci {et~al.}(2011)Pentericci, Fontana, Vanzella, Castellano,
  Grazian, Dijkstra, Boutsia, Cristiani, Dickinson, Giallongo,
  {et~al.}}]{pentericci2011laefrac/z=7LBG}
Pentericci, L., Fontana, A., Vanzella, E., {et~al.} 2011, The Astrophysical
  Journal, 743, 132

\bibitem[{Pentericci {et~al.}(2018)Pentericci, Vanzella, Castellano, Fontana,
  De~Barros, Grazian, Marchi, Bradac, Conselice, Cristiani,
  {et~al.}}]{pentericci2018LAEfrac}
Pentericci, L., Vanzella, E., Castellano, M., {et~al.} 2018, Astronomy \&
  Astrophysics, 619, A147

\bibitem[{Piqueras {et~al.}(2019)Piqueras, Conseil, Shepherd, Bacon, Leclercq,
  Richard, Molinaro, Shortridge, \& Pasian}]{piqueras2019asp}
Piqueras, L., Conseil, S., Shepherd, M., {et~al.} 2019, ASP Conf. Ser. Vol.
  521, Astronomical Data Analysis Software and Systems XXVI

\bibitem[{{Planck Collaboration} {et~al.}(2020){Planck Collaboration},
  {Aghanim, N.}, {Akrami, Y.}, {Ashdown, M.}, {Aumont, J.}, {Baccigalupi, C.},
  {Ballardini, M.}, {Banday, A. J.}, {Barreiro, R. B.}, {Bartolo, N.}, {Basak,
  S.}, {Battye, R.}, {Benabed, K.}, {Bernard, J.-P.}, {Bersanelli, M.},
  {Bielewicz, P.}, {Bock, J. J.}, {Bond, J. R.}, {Borrill, J.}, {Bouchet, F.
  R.}, {Boulanger, F.}, {Bucher, M.}, {Burigana, C.}, {Butler, R. C.},
  {Calabrese, E.}, {Cardoso, J.-F.}, {Carron, J.}, {Challinor, A.}, {Chiang, H.
  C.}, {Chluba, J.}, {Colombo, L. P. L.}, {Combet, C.}, {Contreras, D.},
  {Crill, B. P.}, {Cuttaia, F.}, {de Bernardis, P.}, {de Zotti, G.},
  {Delabrouille, J.}, {Delouis, J.-M.}, {Di Valentino, E.}, {Diego, J. M.},
  {Dor\'e, O.}, {Douspis, M.}, {Ducout, A.}, {Dupac, X.}, {Dusini, S.},
  {Efstathiou, G.}, {Elsner, F.}, {En\ss{}lin, T. A.}, {Eriksen, H. K.},
  {Fantaye, Y.}, {Farhang, M.}, {Fergusson, J.}, {Fernandez-Cobos, R.},
  {Finelli, F.}, {Forastieri, F.}, {Frailis, M.}, {Fraisse, A. A.},
  {Franceschi, E.}, {Frolov, A.}, {Galeotta, S.}, {Galli, S.}, {Ganga, K.},
  {G\'enova-Santos, R. T.}, {Gerbino, M.}, {Ghosh, T.}, {Gonz\'alez-Nuevo, J.},
  {G\'orski, K. M.}, {Gratton, S.}, {Gruppuso, A.}, {Gudmundsson, J. E.},
  {Hamann, J.}, {Handley, W.}, {Hansen, F. K.}, {Herranz, D.}, {Hildebrandt, S.
  R.}, {Hivon, E.}, {Huang, Z.}, {Jaffe, A. H.}, {Jones, W. C.}, {Karakci, A.},
  {Keih\"anen, E.}, {Keskitalo, R.}, {Kiiveri, K.}, {Kim, J.}, {Kisner, T. S.},
  {Knox, L.}, {Krachmalnicoff, N.}, {Kunz, M.}, {Kurki-Suonio, H.}, {Lagache,
  G.}, {Lamarre, J.-M.}, {Lasenby, A.}, {Lattanzi, M.}, {Lawrence, C. R.}, {Le
  Jeune, M.}, {Lemos, P.}, {Lesgourgues, J.}, {Levrier, F.}, {Lewis, A.},
  {Liguori, M.}, {Lilje, P. B.}, {Lilley, M.}, {Lindholm, V.},
  {L\'opez-Caniego, M.}, {Lubin, P. M.}, {Ma, Y.-Z.}, {Mac\'{\i}as-P\'erez, J.
  F.}, {Maggio, G.}, {Maino, D.}, {Mandolesi, N.}, {Mangilli, A.},
  {Marcos-Caballero, A.}, {Maris, M.}, {Martin, P. G.}, {Martinelli, M.},
  {Mart\'{\i}nez-Gonz\'alez, E.}, {Matarrese, S.}, {Mauri, N.}, {McEwen, J.
  D.}, {Meinhold, P. R.}, {Melchiorri, A.}, {Mennella, A.}, {Migliaccio, M.},
  {Millea, M.}, {Mitra, S.}, {Miville-Desch\^enes, M.-A.}, {Molinari, D.},
  {Montier, L.}, {Morgante, G.}, {Moss, A.}, {Natoli, P.},
  {N\o{}rgaard-Nielsen, H. U.}, {Pagano, L.}, {Paoletti, D.}, {Partridge, B.},
  {Patanchon, G.}, {Peiris, H. V.}, {Perrotta, F.}, {Pettorino, V.},
  {Piacentini, F.}, {Polastri, L.}, {Polenta, G.}, {Puget, J.-L.}, {Rachen, J.
  P.}, {Reinecke, M.}, {Remazeilles, M.}, {Renzi, A.}, {Rocha, G.}, {Rosset,
  C.}, {Roudier, G.}, {Rubi\~no-Mart\'{\i}n, J. A.}, {Ruiz-Granados, B.},
  {Salvati, L.}, {Sandri, M.}, {Savelainen, M.}, {Scott, D.}, {Shellard, E. P.
  S.}, {Sirignano, C.}, {Sirri, G.}, {Spencer, L. D.}, {Sunyaev, R.},
  {Suur-Uski, A.-S.}, {Tauber, J. A.}, {Tavagnacco, D.}, {Tenti, M.},
  {Toffolatti, L.}, {Tomasi, M.}, {Trombetti, T.}, {Valenziano, L.},
  {Valiviita, J.}, {Van Tent, B.}, {Vibert, L.}, {Vielva, P.}, {Villa, F.},
  {Vittorio, N.}, {Wandelt, B. D.}, {Wehus, I. K.}, {White, M.}, {White, S. D.
  M.}, {Zacchei, A.}, \& {Zonca, A.}}]{Planck2018reionisation}
{Planck Collaboration}, {Aghanim, N.}, {Akrami, Y.}, {et~al.} 2020, A\&A, 641,
  A6

\bibitem[{Richard {et~al.}(2021)Richard, Claeyssens, Lagattuta, Guaita, Bauer,
  Pello, Carton, Bacon, Soucail, Lyon, {et~al.}}]{richard2021atlas}
Richard, J., Claeyssens, A., Lagattuta, D., {et~al.} 2021, Astronomy \&
  Astrophysics, 646, A83

\bibitem[{Robertson(2022)}]{robertson2022Reionisationreview}
Robertson, B.~E. 2022, Annual Review of Astronomy and Astrophysics, 60, 121

\bibitem[{Robertson {et~al.}(2013)Robertson, Furlanetto, Schneider, Charlot,
  Ellis, Stark, McLure, Dunlop, Koekemoer, Schenker,
  {et~al.}}]{robertson2013reionisation}
Robertson, B.~E., Furlanetto, S.~R., Schneider, E., {et~al.} 2013, The
  Astrophysical Journal, 768, 71

\bibitem[{Salpeter(1955)}]{salpeter1955IMF}
Salpeter, E.~E. 1955, The Astrophysical Journal, 121, 161

\bibitem[{Santos {et~al.}(2020)Santos, Sobral, Matthee, Calhau, Da~Cunha,
  Ribeiro, Paulino-Afonso, Arrabal~Haro, \&
  Butterworth}]{santos2020lyaUVproperties}
Santos, S., Sobral, D., Matthee, J., {et~al.} 2020, Monthly Notices of the
  Royal Astronomical Society, 493, 141

\bibitem[{Schaerer {et~al.}(2011)Schaerer, de~Barros, \&
  Stark}]{schaerer2011LAEfrac}
Schaerer, D., de~Barros, S., \& Stark, D.~P. 2011, Astronomy \& Astrophysics,
  536, A72

\bibitem[{Schenker {et~al.}(2014)Schenker, Ellis, Konidaris, \&
  Stark}]{schenker2014lyafracfeasibility}
Schenker, M.~A., Ellis, R.~S., Konidaris, N.~P., \& Stark, D.~P. 2014, The
  Astrophysical Journal, 795, 20

\bibitem[{Shipley {et~al.}(2018)Shipley, Lange-Vagle, Marchesini, Brammer,
  Ferrarese, Stefanon, Kado-Fong, Whitaker, Oesch, Feinstein,
  {et~al.}}]{shipley2018hff}
Shipley, H.~V., Lange-Vagle, D., Marchesini, D., {et~al.} 2018, The
  Astrophysical Journal Supplement Series, 235, 14

\bibitem[{Smit {et~al.}(2017)Smit, Swinbank, Massey, Richard, Smail, \&
  Kneib}]{smit2017lensedemissionlines}
Smit, R., Swinbank, A., Massey, R., {et~al.} 2017, Monthly Notices of the Royal
  Astronomical Society, 467, 3306

\bibitem[{Smith {et~al.}(2022)Smith, Kannan, Garaldi, Vogelsberger, Pakmor,
  Springel, \& Hernquist}]{smith2022lyatransmission_sim}
Smith, A., Kannan, R., Garaldi, E., {et~al.} 2022, Monthly Notices of the Royal
  Astronomical Society, 512, 3243

\bibitem[{Sobral \& Matthee(2019)}]{sobral2019predictinglyaesc}
Sobral, D. \& Matthee, J. 2019, Astronomy \& Astrophysics, 623, A157

\bibitem[{Stark(2016)}]{Stark2016review}
Stark, D.~P. 2016, Annual Review of Astronomy and Astrophysics, 54, 761

\bibitem[{Stark {et~al.}(2017)Stark, Ellis, Charlot, Chevallard, Tang, Belli,
  Zitrin, Mainali, Gutkin, Vidal-Garc{\'\i}a,
  {et~al.}}]{stark2017luminouslyalpha}
Stark, D.~P., Ellis, R.~S., Charlot, S., {et~al.} 2017, Monthly Notices of the
  Royal Astronomical Society, 464, 469

\bibitem[{Stark {et~al.}(2010)Stark, Ellis, Chiu, Ouchi, \&
  Bunker}]{stark2010keckLAEfrac}
Stark, D.~P., Ellis, R.~S., Chiu, K., Ouchi, M., \& Bunker, A. 2010, Monthly
  Notices of the Royal Astronomical Society, 408, 1628

\bibitem[{Stark {et~al.}(2011)Stark, Ellis, \& Ouchi}]{stark2011LAEfrac}
Stark, D.~P., Ellis, R.~S., \& Ouchi, M. 2011, The Astrophysical Journal
  Letters, 728, L2

\bibitem[{{Thai} {et~al.}(2023){Thai}, {Tuan-Anh}, {Pello}, \&
  {Goovaerts}}]{Thai2023}
{Thai}, T.~T., {Tuan-Anh}, P., {Pello}, R., \& {Goovaerts}, I. 2023, Astronomy
  \& Astrophysics, submitted

\bibitem[{Verhamme {et~al.}(2008)Verhamme, Schaerer, Atek, \&
  Tapken}]{verhamme2008lyadustsim}
Verhamme, A., Schaerer, D., Atek, H., \& Tapken, C. 2008, Astronomy \&
  Astrophysics, 491, 89

\bibitem[{Weilbacher {et~al.}(2020)Weilbacher, Palsa, Streicher, Bacon,
  Urrutia, Wisotzki, Conseil, Husemann, Jarno, Kelz, {et~al.}}]{PW2020MUSEdata}
Weilbacher, P.~M., Palsa, R., Streicher, O., {et~al.} 2020, Astronomy \&
  Astrophysics, 641, A28

\bibitem[{Wisotzki {et~al.}(2016)Wisotzki, Bacon, Blaizot, Brinchmann, Herenz,
  Schaye, Bouch{\'e}, Cantalupo, Contini, Carollo,
  {et~al.}}]{wisotzki2016extendedlyahaloes}
Wisotzki, L., Bacon, R., Blaizot, J., {et~al.} 2016, Astronomy \& Astrophysics,
  587, A98

\bibitem[{Witten {et~al.}(2023)Witten, Laporte, Martin-Alvarez, Sijacki, Yuan,
  Haehnelt, Baker, Dunlop, Ellis, Grogin, Illingworth, Katz, Koekemoer, Magee,
  Maiolino, McClymont, Pérez-González, Puskas, Roberts-Borsani, Santini, \&
  Simmonds}]{Witten2023lya_inreionisation}
Witten, C., Laporte, N., Martin-Alvarez, S., {et~al.} 2023, arXiv preprint
  arXiv:2303.16225v1

\bibitem[{Yoshioka {et~al.}(2022)Yoshioka, Kashikawa, Inoue, Yamanaka,
  Shimasaku, Harikane, Shibuya, Momose, Ito, Liang,
  {et~al.}}]{yoshioka2022LAEfrac}
Yoshioka, T., Kashikawa, N., Inoue, A.~K., {et~al.} 2022, The Astrophysical
  Journal, 927, 32

\bibitem[{Zheng {et~al.}(2010)Zheng, Cen, Trac, \&
  Miralda-Escud{\'e}}]{zheng2010lya_radiativetransfer}
Zheng, Z., Cen, R., Trac, H., \& Miralda-Escud{\'e}, J. 2010, The Astrophysical
  Journal, 716, 574

\end{thebibliography}
\end{document}